\documentclass{jfm}
\usepackage{epsfig}
\usepackage{natbib}
\usepackage{color}
\usepackage{amsmath}
\usepackage{amssymb}
\usepackage{upmath}

\ifCUPmtlplainloaded \else
  \checkfont{eurm10}
  \iffontfound
    \IfFileExists{upmath.sty}
      {\typeout{^^JFound AMS Euler Roman fonts on the system,
                   using the 'upmath' package.^^J}%
       \usepackage{upmath}}
      {\typeout{^^JFound AMS Euler Roman fonts on the system, but you
                   dont seem to have the}%
       \typeout{'upmath' package installed. JFM.cls can take advantage
                 of these fonts,^^Jif you use 'upmath' package.^^J}%
      }
  \else
  \fi
\fi

\ifCUPmtlplainloaded \else
  \checkfont{msam10}
  \iffontfound
    \IfFileExists{amssymb.sty}
      {\typeout{^^JFound AMS Symbol fonts on the system, using the
                'amssymb' package.^^J}%
       \usepackage{amssymb}%
         \let\leq=\leqslant
         
      }{}
  \fi
\fi

\ifCUPmtlplainloaded \else
  \IfFileExists{amsbsy.sty}
    {\typeout{^^JFound the 'amsbsy' package on the system, using it.^^J}%
     \usepackage{amsbsy}}
    {\providecommand\boldsymbol[1]{\mbox{\boldmath $##1$}}}
\fi

\newcommand{\Ca}{\mbox{\textit{Ca}}} % Froude number

\newsavebox{\astrutbox}
\sbox{\astrutbox}{\rule[-5pt]{0pt}{20pt}}

\title[Global stability of stretched jets]{Global stability of stretched
jets: conditions for the generation of monodisperse micro-emulsions using coflows}

\author[J.M. Gordillo, A. Sevilla and F. Campo-Cort\'es]
{J.\ns M.\ns G\ls O\ls R\ls D\ls I\ls L\ls L\ls O\ls$^{1}$, \ns
A.\ns S\ls E\ls V\ls I\ls L\ls L\ls A\ls$^{2}$, \ns F.\ns C\ls A\ls M\ls P\ls O\ls-\ls C\ls O\ls
R\ls T\ls\'E\ls S\ls$^{1}$}
\affiliation{$^{1}$\'Area de Mec\'anica de Fluidos, Departamento de Ingener\'ia
Aeroespacial y Mec\'anica de Fluidos, Universidad de Sevilla,
Avenida de los Descubrimientos s/n 41092, Sevilla, Spain\\ $^{2}$ \'Area de
Mec\'anica de Fluidos, Departamento de Ingenier\'ia T\'ermica y de Fluidos.
Universidad Carlos III de Madrid. Avda. de la Universidad 30, 28911, Legan\'es, Spain.}

\date{Received 24 May 2013; revised 10 September 2013; accepted 4 November 2013}

\begin{document}

\maketitle

\begin{abstract}
In this paper we reveal the physics underlaying the conditions needed for the generation
of emulsions composed of uniformly sized drops of micrometric or submicrometric diameters
when two immiscible streams flow in parallel under the so-called tip streaming regime
after~\cite*{SuryoBasaran}. Indeed, when inertial effects in both liquid streams are
negligible, the inner to outer flow-rate and viscosity ratios are small enough and the capillary number
is above an experimentally determined threshold which is predicted by our theoretical
results with small relative errors, a steady micron-sized jet is issued from the apex
of a conical drop. Under these conditions, the jet disintegrates into drops with a very
well defined mean diameter, giving rise to a monodisperse microemulsion. Here, we
demonstrate that the regime in which uniformly-sized drops are produced corresponds to
values of the capillary number for which the cone-jet system is \emph{globally stable}.
Interestingly enough, our general stability theory reveals that liquid jets with a
cone-jet structure are much more stable than their cylindrical counterparts thanks, mostly,
to a capillary stabilization mechanism described here for the first
time. Our findings also limit the validity of the type of stability analysis based on
the common parallel flow assumption to only those situations in which the liquid jet
diameter is almost constant.
\end{abstract}

% \begin{keywords}
% Stretched jets, global instability, viscous co-flows, micro-emulsions.
% \end{keywords}

%%%%%%%%%%%%%%%%%%%%%%%%%%%%%%%%%%%%%%%%%%%%%%%%%%%%%%
%%%%%%%%%%%%%%%%% Introduction %%%%%%%%%%%%%%%%%%%%%%%
%%%%%%%%%%%%%%%%%%%%%%%%%%%%%%%%%%%%%%%%%%%%%%%%%%%%%%

\section{Introduction}
\label{sec:introduction}

The controlled production of emulsions composed of micron or submicron-sized drops with a
well defined mean size possesses uncountable applications in industry, medicine and
pharmacology~\citep{Basaranaiche,AnnuRevStone,BarreroAnnurev}. One of the most successful
methodologies for the production of micro or nano drops consists in generating a thin
thread of the fluid to be dispersed within a coflowing stream of the carrier one. Under
these conditions, the narrow jet subsequently breaks downstream of the injection tube due
to the growth of capillary disturbances. To avoid the clogging of these type of
flow-focusing~\citep*{Anna} or coflow devices~\citep*{SuryoBasaran,UtadaPRL,Marin}, the
diameter of the feeding tube is usually much larger than that of the produced jet which,
therefore, suffers a strong stretching in the downstream direction.

\begin{figure}
\centering
\includegraphics[width=\textwidth]{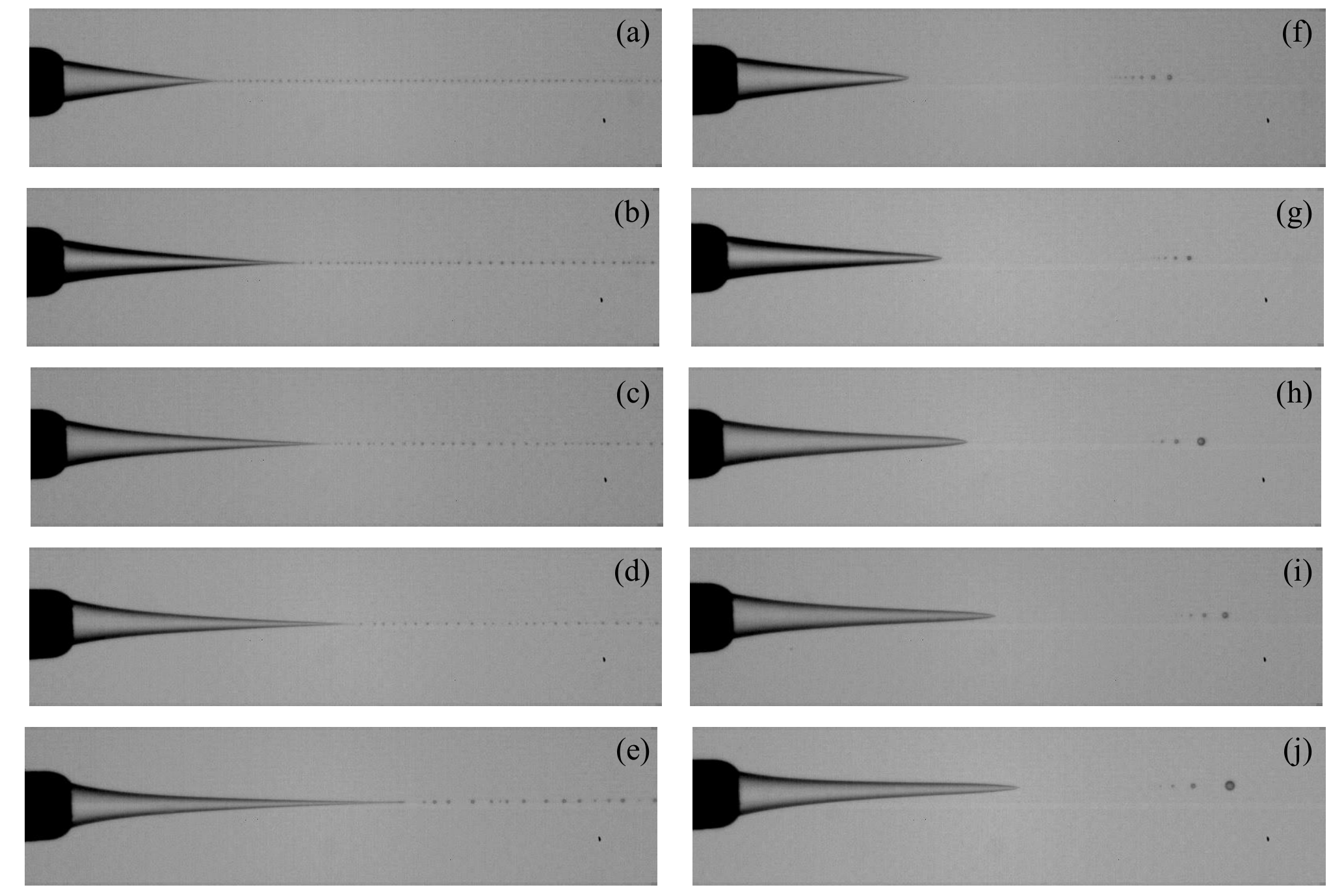}
\caption{For each of the values of the viscosity ratio considered in this study, namely,
(from top to bottom) $\lambda=10^{-2}$, $\lambda=5\times 10^{-3}$, $\lambda=3\times 10^{-3}$,
$\lambda=2\times 10^{-3}$ and $\lambda=10^{-3}$, there exists a critical value of the outer
velocity $U_o$ and, consequently, of the capillary number $\Ca=\mu_o\,U_o/\sigma$, above which
a steady jet issues from the apex of a conical drop and a train of uniform-sized drops is
produced (left column, figures 1a--e). However, if the capillary number is below its critical
value, the jet issued at the apex of the cone is unsteady and breaks unevenly, forming drops
of different sizes (right column, figures 1f--j). The values of $\Ca$ and $\lambda$ in each of
the experimental images depicted in this figure are the following:
(a) $\lambda=10^{-2}$, $\Ca=1.12$, (b) $\lambda=5\times10^{-3}$, $\Ca=1.15$,
(c) $\lambda=3\times 10^{-3}$, $\Ca=1.7$, (d) $\lambda=2\times 10^{-3}$, $\Ca=2.09$,
(e) $\lambda=10^{-3}$, $\Ca=2.69$, (f) $\lambda=10^{-2}$, $\Ca=0.9$,
(g) $\lambda=5\times10^{-3}$, $\Ca=0.98$, (h) $\lambda=3\times 10^{-3}$, $\Ca=1.37$,
(i) $\lambda=2\times 10^{-3}$, $\Ca=1.69$, (j) $\lambda=10^{-3}$, $\Ca=2.06$.
The outer diameter of the injection tube is $200$ $\mu$m.\label{fig1}}
\end{figure}

\begin{figure}
\centering
\includegraphics[width=\textwidth]{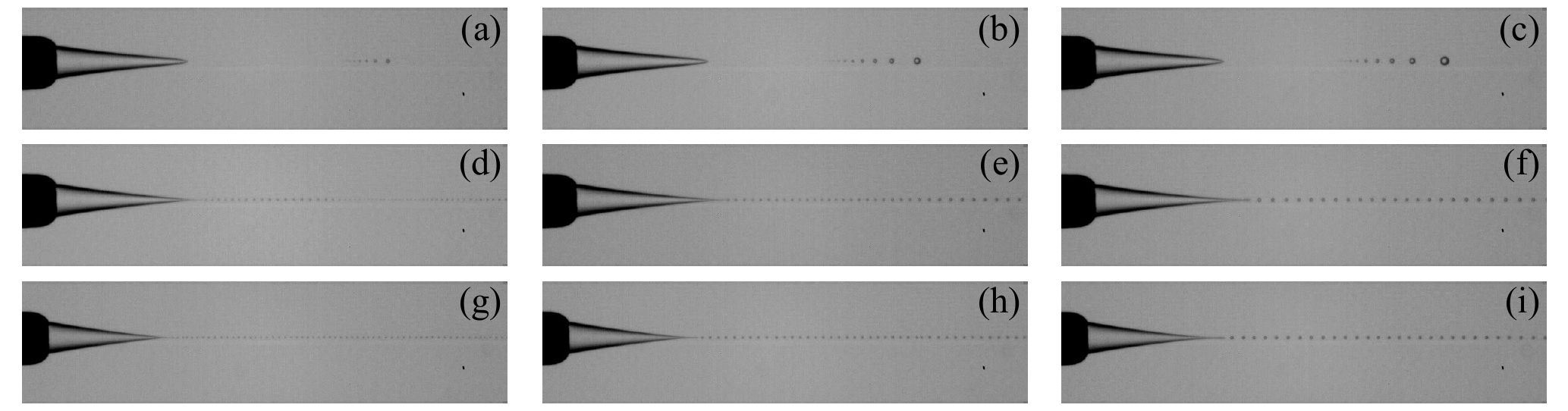}
\caption{The figure shows the effect of varying the flow rate ratio $q$, which increases
from left to right, on the drop formation processes for a fixed value of the viscosity
ratio, $\lambda=10^{-2}$, and three values of the capillary number, namely, $\Ca=0.9$
(top row), $\Ca=0.98$ (middle row) and $\Ca=1.12$ (bottom row). The analysis of the images
reveals that the value of the critical capillary number, which in this case is between
$\Ca=0.90$ and $\Ca=1.12$ is rather insensitive to changes in $q$. The same conclusion
can be extracted from the analysis of the experimental images corresponding to the rest
of viscosity ratios investigated. The values of $Q_i$ in each of the experimental images
are the following: a) $Q_i=2.1\times10^{-4}$ mm$^3\,$s$^{-1}$, b) $Q_i=3.8\times10^{-3}$
 mm$^3\,$s$^{-1}$, c) $Q_i=8.6\times10^{-3}$ mm$^3\,$s$^{-1}$, d) $Q_i=7.3\times10^{-4}$
 mm$^3\,$s$^{-1}$, e) $Q_i=4.2\times10^{-3}$ mm$^3\,$s$^{-1}$, f) $Q_i=7.4\times10^{-3}$
 mm$^3\,$s$^{-1}$, g) $Q_i=1.3\times10^{-3}$ mm$^3\,$s$^{-1}$, h) $Q_i=2.9\times10^{-3}$
 mm$^3\,$s$^{-1}$, i) $Q_i=1.4\times10^{-2}$ mm$^3\,$s$^{-1}$.
The outer diameter of the injection tube is $200$ $\mu$m \label{fig2}}
\end{figure}

Recently,~\citet*{JFM12} reported experiments on the generation of concentrated monodisperse
emulsions composed of drops with \emph{uniform} sizes of even below 1 $\mu$m using a low
Reynolds number co-flow configuration inspired by the numerical experiments due
to~\cite{SuryoBasaran}. In these experiments a flow rate $Q_i$ of a fluid with a viscosity
$\mu_i$ discharges through a cylindrical tube of inner radius $R_i$ into an immiscible liquid
of viscosity $\mu_o=\mu_i/\lambda$ flowing in parallel with the axis of the injector at a
velocity $U_o$ with, from now on, $\sigma$ indicating the interfacial tension coefficient
between both immiscible liquids and $\lambda$ the inner-to-outer viscosity ratio. The
experimental observations of figures~\ref{fig1} and~\ref{fig2} reveal that the inner jet
highly stretches downstream provided that the flow-rate ratio $q=Q_i/(\pi\,R^2_iU_o)$ is
sufficiently small and the capillary number $\Ca=\mu_o\,U_o/\sigma$ is of order unity or
larger. The approximately cylindrical ligament emitted from the tip of the conical drop
breaks under the action of capillary forces, giving rise to a train of droplets with a low
size dispersion \emph{if} $\Ca$ is sufficiently large. Under these conditions, it was found
in~\citet{JFM12} that the drop diameter scales as $D_d \propto R_i q^{1/2}$. Therefore,
uniformly-sized drops with arbitrarily small diameters could be, in principle, obtained if
the control parameter $q$ is set to sufficiently small values and the capillary number
$\Ca$ is above a so far unknown threshold capillary number $\Ca^*$, whose determination
as a function of $q$ and $\lambda$, with $\lambda\ll 1$, is the main purpose of the present work.

To illustrate the effect of the capillary number on the distribution of drop sizes,
figure~\ref{fig1} as well as the three movies provided as supplementary material, show how water drops
are generated and dispersed within co-flowing
streams of different silicon oils with viscosities ranging from $\mu_o=100$ cp to
$\mu_o=1000$ cp at $25^o$$C$. Figures~\ref{fig1}a-e reveal that, for each value of the
viscosity ratio investigated, there exists a threshold outer velocity $U_o$ above which
the thin liquid thread ejected from the tip of a conical drop breaks regularly, forming
tiny droplets with very similar diameters. In the experimental situations depicted in
figures~\ref{fig1}f-j, the overall geometry of the drop attached at the injection tube
is similar to that showed in figures~\ref{fig1}a-e, but the jet issued at the apex of
the cone is unsteady and breaks unevenly, forming droplets with a broad size distribution.
It is interesting to note that the cone itself is steady even in cases in which the drop
formation process is not periodic.

The reason for the differences observed in the two series of experiments depicted in
figure~\ref{fig1} is that, in figures~\ref{fig1}a-e $\Ca>\Ca^*$, whereas $\Ca<\Ca^*$ in
figures~\ref{fig1}f-j, where $\Ca^*$ is defined as the minimum value of the capillary number
for which a train of uniform-sized drops is produced for fixed values of $\lambda$ and $q$.
It is also interesting to note that, in contrast with the noticeable dependence of $\Ca^*$
on $\lambda$ deduced from the analysis of the type of experimental images depicted in
figure~\ref{fig1}, the results in figure~\ref{fig2} reveal that the critical capillary number
is rather insensitive to changes in $q$.

\begin{figure}
\centering
\includegraphics[width=1\textwidth]{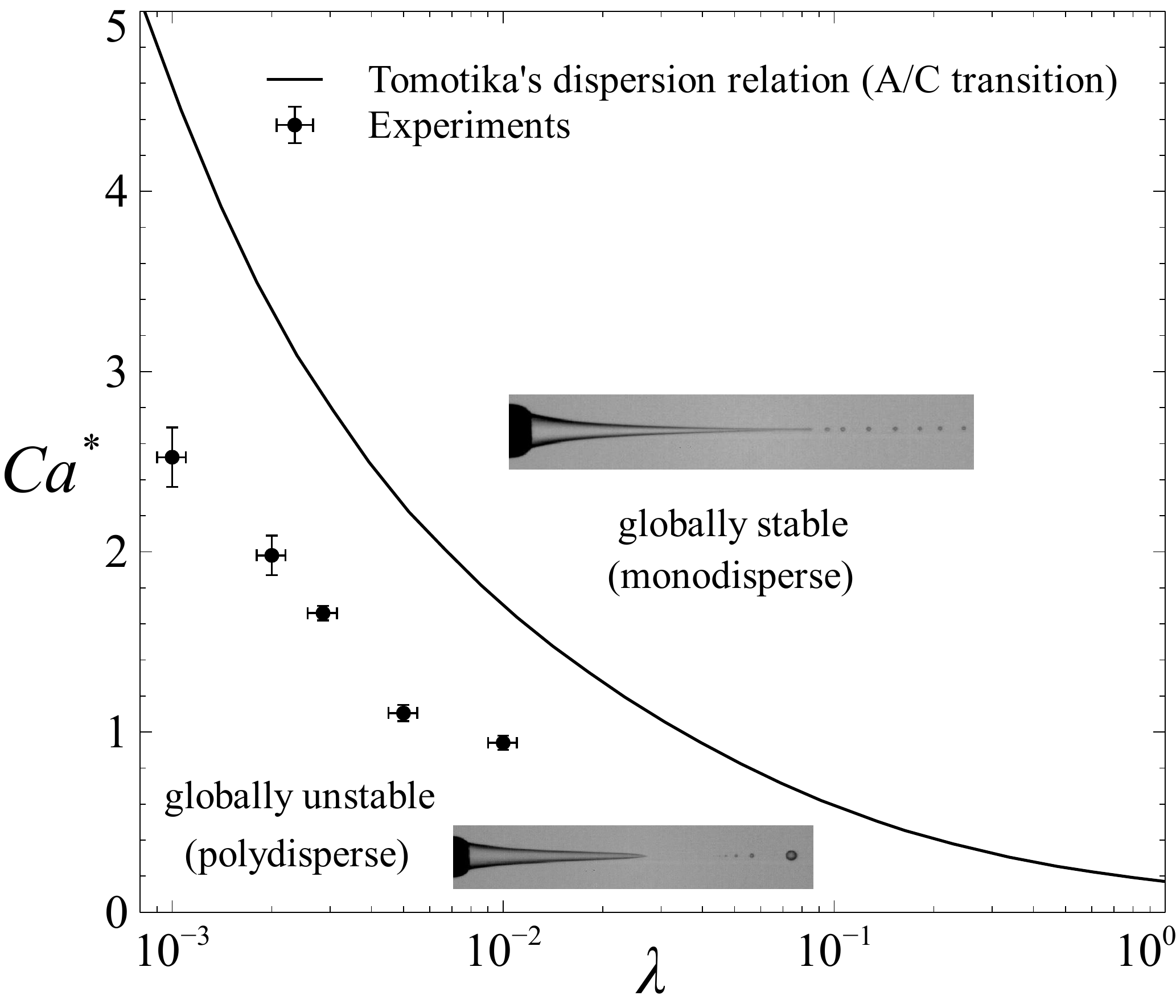}
\caption{The critical capillary number, $\Ca^*$, as a function of the viscosity ratio $\lambda$.
Symbols show the experimental transition points separating the monodisperse ($\Ca>\Ca^*$,
figures 1a--e), and polydisperse ($\Ca<\Ca^*$, figures 1f--j) droplet generation regimes.
The line shows the critical capillary number for the transition from convective to absolute
instability computed using Tomotika's dispersion relation.
\label{fig3}}
\end{figure}

Since many different applications such as drug delivery demand that the size distribution of
the drops composing the emulsion is as narrow as possible, it is desirable that the capillary
number at which the coflow device is operated is larger than $\Ca^*$. It is thus our purpose in
this paper to develop a theory to quantify and explain why drops with uniform sizes are produced
only for values of the capillary number above the critical value $\Ca^*$ deduced from the analysis of the experimental images in figures~\ref{fig1} and~\ref{fig2}. Notice that, according to the experimental
observations in figures~\ref{fig1} and~\ref{fig2}, figure~\ref{fig3} shows that
$\Ca^*$ is a decreasing function of $\lambda$ which does not appreciably vary with $q$.

As a first attempt to predict and quantify the dependence of the critical capillary number
on $\lambda$ depicted in figure~\ref{fig3}, let us recall that the non-regularity in the drop
formation process is associated with the fact that the thin liquid thread from which drops are
emitted, is itself unsteady. This experimental evidence suggests that the unstable capillary
perturbations that give rise to the generation of drops cannot be convected downstream at a
sufficiently large speed when $\Ca<\Ca^*$. Therefore, the two regimes delimited by the
experimental data points in figure~\ref{fig3} should correspond to those conditions for which
unstable perturbations are either convected downstream the cone tip (convectively unstable
flow), or are able to propagate upstream, thus inhibiting the formation of a steady jet
(absolutely unstable flow).

Previous studies~\citep{PoF05,UtadaPRL,UtadaPRL08,GuillotPRL} have demonstrated that the experimentally observed transition from jetting to dripping in cylindrical capillary jets can be predicted by means of a \emph{local} linear stability
analysis of parallel streams~\citep{AnnRevHuerre}. Thus, with the purpose of calculating the
boundary separating the two different drop formation processes depicted in figures~\ref{fig1}, \ref{fig2} and in the movies provided as supplementary material, we have determined the values of the capillary number for which the temporal growth rate of the perturbations with zero
group velocity is also equal to zero. The result obtained using Tomotika's dispersion
relation~\citep{Tomotika,PRLGoldstein,PoFPowers,PREInnombrable}, which describes the growth and
propagation of perturbations in a cylindrical jet immersed into another immiscible liquid,
is represented together with the experimental data in figure~\ref{fig3}. Although the
conditions of validity of Tomotika's analysis are fulfilled at large distances from the
injector, where the diameter of the liquid jet is nearly constant, figure~\ref{fig3} reveals
that the parallel flow stability analysis predicts larger values for the critical capillary
number than those measured experimentally. Consequently, the full jet, which strongly stretches
in the downstream direction, is more stable than the cylindrical portion located far downstream.

To improve the agreement with experimental results, in this study we have performed a
\emph{global stability analysis} that takes into account the real shape of the stretched jet
and describes the growth and propagation of perturbations without resorting to the parallel
flow simplification used in previous studies~\citep{Tomotika,PREInnombrable}. To avoid the
lengthy numerical computations associated with a global stability analysis in two or three
spatial directions~\citep[see e.g.][]{Estela1,Christodoulou,Estela2,Theofilis2011}, in the present work
we have developed a one-dimensional model based on slender-body theory that yields a single
partial differential equation governing the spatiotemporal evolution of the jet radius. In \cite{JFM12} it was shown that
the axial variations of both the steady shape of the jet and the outer flow field are very well
reproduced by our previous theory. Here we will demonstrate that our global stability analysis, that resorts to a new equation for the jet radius, faithfully reproduces Tomotika's dispersion relation in the long wave limit for the case of cylindrical jets of small viscosity ratio. Moreover, the experimentally measured values of the critical
capillary number are very well reproduced since our final equation retains the axial dependence of the flow field in the formulation. Let us point out that the global stability analysis, developed in \S 2.2, shares many
similarities with that by~\citet*{Rubio2013}, where the global stability of gravitationally
stretched jets is studied using a different one-dimensional model and excellent agreement with experiments
is found.

\section{Unsteady slender-body theory for the description of highly stretched jets in Stokes flow}
\label{sec:theory}
\begin{figure}
\begin{center}
\includegraphics[width=1\columnwidth]{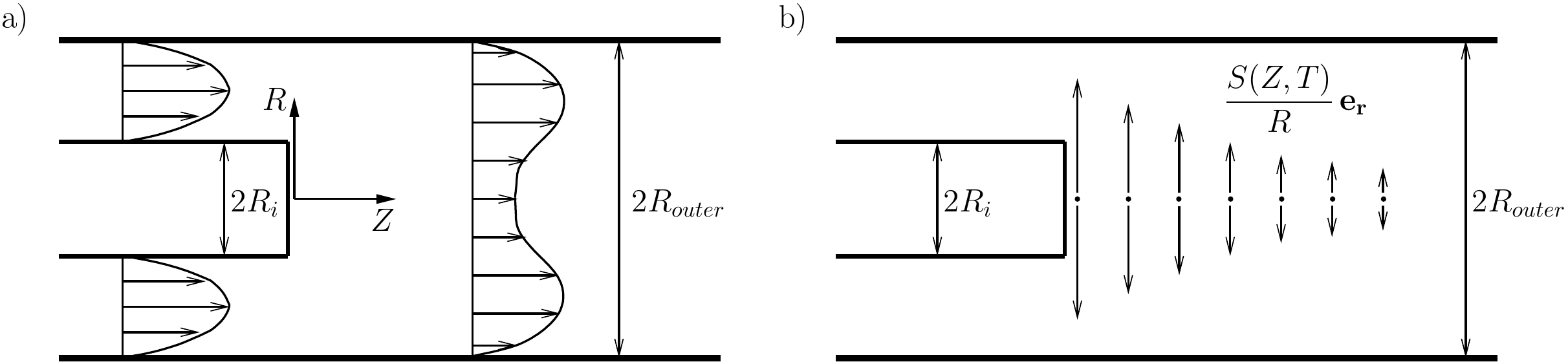}
\end{center}
\caption{Sketch showing that the outer velocity field is decomposed
as the sum of two simpler velocity fields: (a) the one that does not take
into account the presence of the inner fluid and (b) the flow field
induced by the jet, which can be approximated as a line of
sources of intensity $S(Z,T)$ located at the axis of symmetry.}\label{exterior}
\end{figure}

One of the possible ways to analyze the conditions under which drops are formed regularly from the tip of the conical drop, would be to solve the unsteady velocity field using the integral formulation due to \cite{Laz}, as was done in ~\citet*{JFM12}. However, following a strategy similar to that successfully applied to the case of liquid jets in air by \cite{GyC,EggersDupont,basaran04b,Rubio2013}, the dynamics of the low Reynolds number jets depicted in figures 1-2 can be described using a much simpler one-dimensional approach which reveals the physics underlying the transition observed in these figures in a neat way. The central idea behind our one-dimensional theory can be understood by noticing that the slender jets shown in figures 1-2 resemble the case of drops immersed in a purely straining flow \cite{Stone94}. The deformation of slender drops immersed in this type of flow field has been rigorously analyzed using asymptotic methods by \cite{Buckmaster,AcrivosyLo} and, from these studies, it is learned 
that the outer flow field in the case of low viscous drops can be decomposed, in a first approximation, as the addition of the unperturbed flow field plus a distribution of sources located at the axis of symmetry. Since, in our case, the capillary that confines the outer fluid is ten times larger than the inner one, the dynamics of the jet is not affected by this confinement, as already demonstrated in \cite{JFM12}. Therefore, the conclusions derived from \cite{Buckmaster,AcrivosyLo} are fully applicable to the situation at hand, and figure 4 sketches the key idea under which our theory is built. This figure shows that the outer velocity field can be decomposed, in a first approximation, as the addition of two simpler flow fields: the unperturbed velocity field, which is the one that would exist in the coflowing device \emph{if the inner fluid was not injected} (see
figure \ref{exterior}a) and that induced on the outer stream by the presence of the jet (see \ref{exterior}b), which can be viewed as a perturbation to the former. From now on, $T$, $R$ and $Z$ will indicate time and the radial and axial coordinates respectively, with $Z=0$ located at the exit of the injection tube, whereas $R_j(Z,T)$ will denote the radius of the jet. In our approach, the unperturbed velocity field is calculated numerically, by solving the Navier-Stokes equations in the zero Reynolds number limit (Stokes equations) subjected to the appropriate boundary conditions, as it will be explained below. Regarding the perturbed flow, we will take advantage of the fact that the outer velocity field induced by a slender jet of a low viscosity fluid immersed in an outer axisymmetric flow, as is the case of the experiments shown in figures 1 and 2, can be approximated to that created by a line of sources located at the axis of symmetry \citep{Ashley,Taylor1964,Buckmaster}. The intensity of the source 
distribution, $S(Z,T)$, must be such that the kinematic boundary condition is verified at the jet interface (see figure \ref{exterior}b).

\begin{figure}
\begin{center}
\includegraphics[width=0.45\columnwidth]{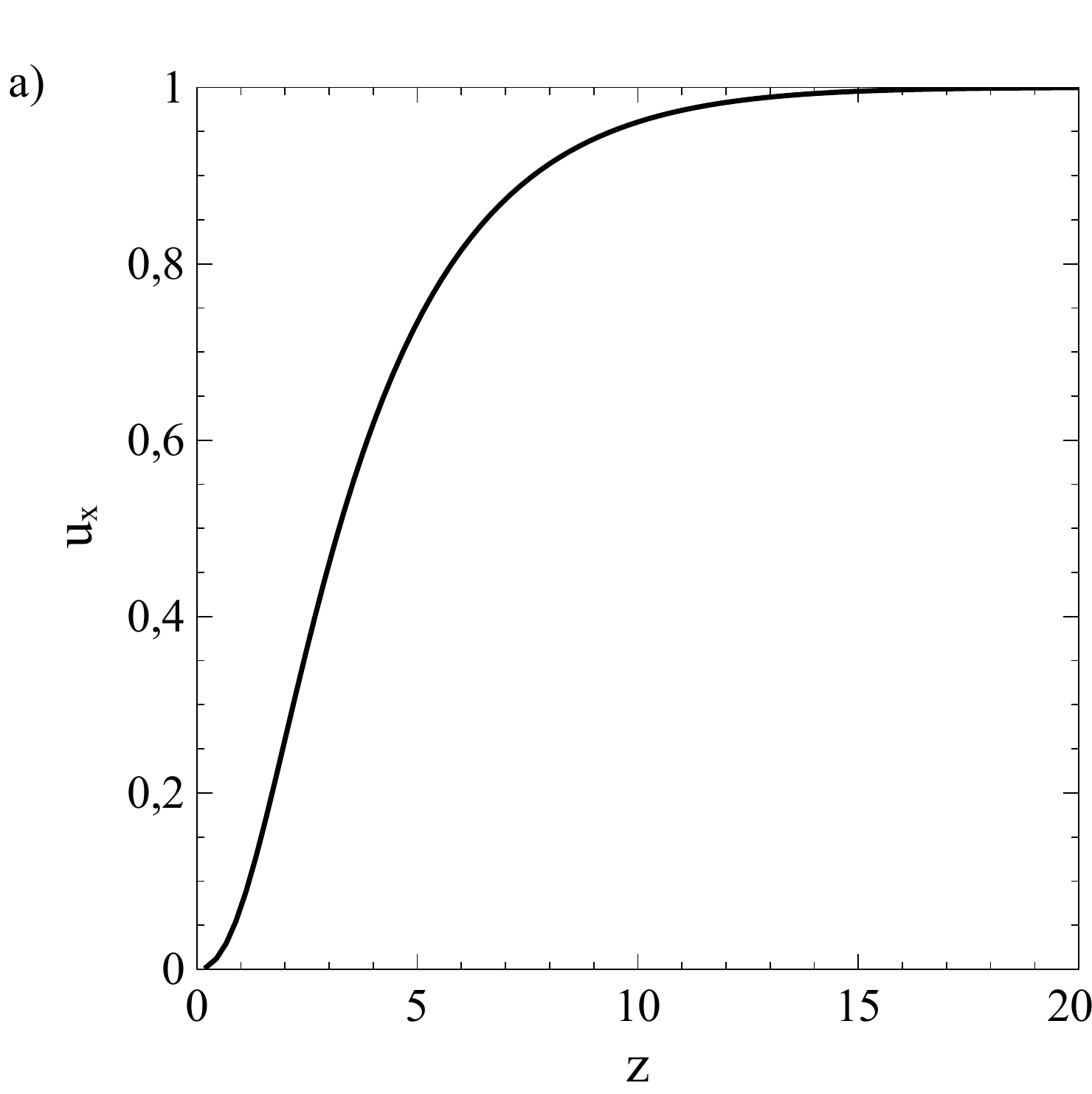}
\includegraphics[width=0.45\columnwidth]{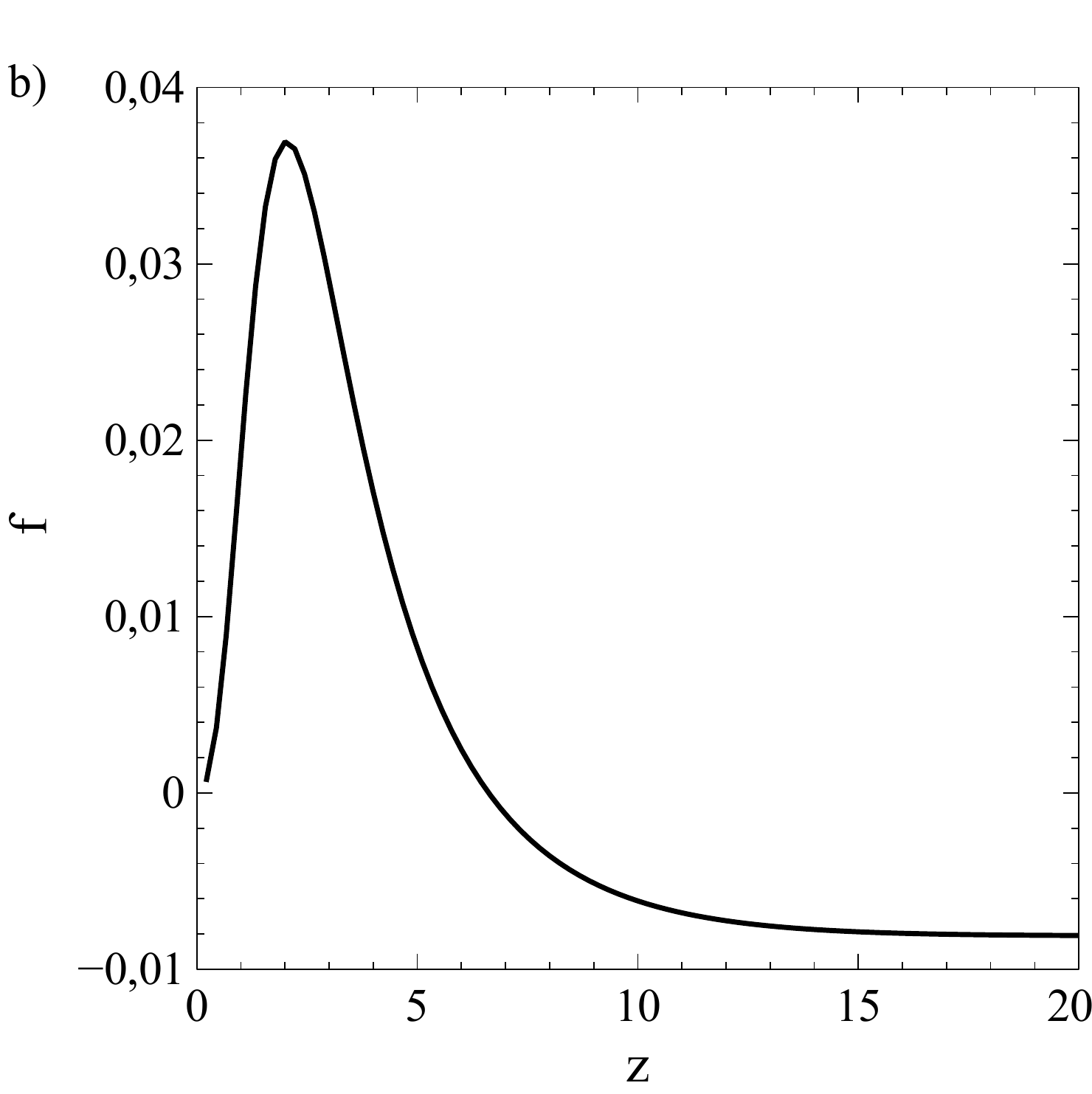}
\end{center}
\caption{Downstream variation of the velocity at the axis of
symmetry, $u_x$ (a) and  $f=1/2\,\partial^2\, u/\partial r^2 (r=0)$ (b)
corresponding to the situation sketched in figure \ref{exterior}a,
i.e, when the inner jet is not present. The dimensionless counterparts of $U_x$ and $F$, which are represented using lower case variables, have been non-dimensionalized using $U_o$ and $R_i$ as characteristic scales of velocity and length.}\label{Uaxis}
\end{figure}

The representation of the outer flow field as the addition of two simpler velocity fields greatly facilitates the theoretical analysis of the spatiotemporal evolution of the jet. Firstly, notice that the numerical computation of the unperturbed flow field sketched in figure \ref{exterior}a is straightforward since this solution does not depend on $Ca$, $\lambda$ or $q$: it is just a function of the geometry of the coflowing device and thus, it is unique for a given injector. But, what's more, notice that the unperturbed velocity field does not even need to be known everywhere in the flow domain; instead, for our purposes, an accurate representation of the flow field in the vicinity of the axis of symmetry is enough. This is due to the fact that our interest here is to deduce an equation for $R_j$, with $R_j/R_i\ll 1$ for $Z/R_i\gtrsim O(1)$. Since the jet radius is located very near the axis of symmetry in most of the flow domain, all the information needed from the numerics to build our theoretical approach 
are just two functions: the axial component of the velocity at the axis, $U_x=U_n(Z,R=0)$, as well as $F(Z)=1/2\, \partial^2 U_n/\partial R^2 (Z,R=0)$, where the subscript $n$ will be used in the following to denote quantities related to the unperturbed problem. Indeed, given these two functions, the analytical expression of the axial component of the unperturbed velocity field in the near-axis region will be expressed in Taylor series as $U_n(Z,R)\simeq U_x(Z)+F(Z)\,R^2+O[(R_j/R_i)^4]$. Moreover, the corresponding approximate expressions of both the radial component of the unperturbed velocity field namely, $V_n(Z,R)$ and of the unperturbed pressure field, $P_n(Z,R)$, will be written in terms of $U_x$ and $F$ making use of the equations of continuity and momentum, as it will be shown below. Let us point out that the main difference between this work and those by \cite{Taylor1964,AcrivosyLo,HinchAcrivos,Sherwood,WendyPRL}, where the deformation of drops and bubbles immersed in a purely straining flow in the 
limit of negligible inertial effects is studied, is that our theory retains the effect of the real outer flow field on the spatiotemporal evolution of the jet. More precisely, the effect of the injector geometry on the outer flow field will be represented in the equation to be deduced for $R_j(Z,T)$ through the two numerically determined functions $U_x(Z)$ and $F(Z)$ which, for the case of the specific geometry used in the experiments explained above, are represented graphically in figure \ref{Uaxis} (see also \cite{JFM12}). Let us point out that the theory to be developed in \S 2.1 can be extended to any other types of axisymmetric geometries provided that the functions $U_x$ and $F$ are straightforwardly determined from the numerical solution of the Stokes equations subjected to the no slip boundary conditions imposed by the geometry of the injector and to the corresponding inflow/outflow boundary conditions.

\subsection{Theory}

The analysis presented in this section generalizes the work by~\citet{JFM12}, and aims to describe
the propagation and growth of capillary disturbances in unsteady coflowing streams with negligible
inertial effects. In our theory, the flow is represented as the superposition of the velocity field created by a distribution of sources with an unknown intensity $S(Z,T)$ plus the unperturbed flow field, which is the solution of the Stokes equations subjected to the the impermeability condition ($\mathbf{U_n}=0$) that needs to be satisfied at:  i) the tube exit, namely, the circle defined by $Z=0$, $R\leq R_i$, ii) the outer surface of the injection tube, which is the cylinder of radius $R=R_i$ that extends along $Z\leq 0$ and iii) the inner surface of the outer tube, which is the cylinder of radius $R=R_{outer}$, where $R_{outer}=10 R_i$ is defined in figure 4. Moreover, Poiseuille velocity profiles are imposed at the upstream and downstream boundaries of the flow domain (see \citet{JFM12} for further details). Regarding the perturbed flow, it was demonstrated by \cite{Taylor1964,AcrivosyLo,HinchAcrivos} that, when a slender drop or bubble with a viscosity much smaller than that of the outer fluid ($\
lambda\ll 1$) is immersed within a straining flow, the first order modification to the unperturbed outer velocity field can be expressed as a superposition of sources located at the axis of symmetry. This is so because the contribution to the perturbed flow field associated with the addition of Stokeslets at $R=0$, is of the order of $\sim O[(R_i/L)^2]$ \citep{Buckmaster}. Here, $L$ denotes the characteristic axial length scale for which the radius of the thread suffers variations of its same order of magnitude and, in the case of slender flows, $R_i/L\ll 1$. The same idea can thus be applied to describe the experiments depicted in figures \ref{fig1}-\ref{fig2} since, in all the cases under study here, $\lambda\ll 1$ and the resulting jet is slender, namely, $\partial\,R_j/\partial\,Z\sim R_i/L\ll 1$. Therefore, the outer flow field can be approximately expressed as~\citep{JFM12},
\begin{equation}
{\bf U}\simeq U_n{\bf
e_z}+\left(V_n+\dfrac{S(Z,T)}{R}\right){\bf
e_r}\simeq\left[U_x(Z)+F(Z) R^2\right]{\bf e_z}+
\left[-\dfrac{R}{2} \dfrac{{\rm d}U_x}{{\rm d}Z}-
\dfrac{R^3}{4}\dfrac{{\rm d}F}{{\rm d}Z}+\dfrac{S(Z,T)}{R}\right]{\bf e_r}\,.\label{U1}
\end{equation}
In equation~\eqref{U1}, $\mathbf{e_r}$ and $\mathbf{e_z}$ are the
unit base vectors,
\begin{equation}
V_n=-\dfrac{R}{2} \dfrac{{\rm d}U_x}{{\rm d}Z}-
\dfrac{R^3}{4}\dfrac{{\rm d}F}{{\rm d}Z}\label{Vn}
\end{equation}
has been expressed as a function of $U_n$ making use of the continuity equation in cylindrical
coordinates, namely, $\partial U_n/\partial Z+R^{-1}\partial (RV_n)/\partial R=0$ and, as it was pointed out above, the axial component of the velocity field is expressed retaining just the first two terms in the in Taylor series expansion of $U_n$, what leads to
\begin{equation}
U_n(R,Z)=U_n(Z,R=0)+1/2\, \partial^2 U_n/\partial\,R^2 (Z,R=0)\,R^2+O[(R_j/R_i)^4]\simeq U_x(Z)+F(Z)\,R^2\, .\label{Un}
\end{equation}
Moreover, in equation (\ref{U1}), only the radial component of the perturbed velocity field created by a distribution of sources, $S(Z,T)/R\,\mathbf{e_r}\sim O(R_j/L)$, has been retained in the analysis due to the fact that the order of magnitude of the corresponding axial component is $\sim O[(R_j/L)^2]\ll 1$ (see, e.g. \cite{Ashley}).

Now notice that the unknown source intensity $S(Z,T)$ can be easily expressed as a function of the jet radius $R_j(Z,T)$ by means of the kinematic boundary condition at the interface,
\begin{equation}
\dfrac{\partial R_j}{\partial T}+\left[U_x(Z)+F(Z)R_j^2(Z,T)\right]
\dfrac{\partial R_j}{\partial Z}=V_n(R=R_j(Z,T))+\dfrac{S(Z,T)}{R_j(Z,T)}.\label{S0}
\end{equation}
Using the expression for $V_n$ given in equation (\ref{Vn}), equation~\eqref{S0} yields
\begin{equation}
\dfrac{S}{R_j}=\dfrac{\partial R_j}{\partial T}+\left(U_x+F R_j^2\right)
\dfrac{\partial R_j}{\partial Z}+\dfrac{R_j}{2}
\dfrac{{\rm d}U_x}{{\rm d}Z}+\dfrac{R_j^3}{4}\dfrac{{\rm d}F}{{\rm d}Z}\,. \label{S}
\end{equation}

Making use of equation (\ref{S}), the equation for $R_j$ can be easily deduced using the continuity equation
\begin{equation}
\dfrac{\partial R^2_j}{\partial T}+\dfrac{\partial}{\partial\,Z}
\left[\left(U_x+F R_j^2\right)R_j^2-\frac{R_j^4}{8\mu_i}
\frac{\partial P_i}{\partial Z}\right]=0\,,\label{Continuidad}
\end{equation}
where it has been taken into account that the flow in the jet is the addition of the plug flow induced by the outer axial velocity field at the radial position where the interface is located plus the Poiseuille flow created by the inner pressure gradient, $-\partial P_i/\partial Z$. It is of interest to point out that the flow within the jet in our one-dimensional theoretical approach may exhibit a recirculation bubble, in contrast with the type of one dimensional models which assume that the velocity profile within the jet is uniform (see, e.g. \cite{Basaran1D2D,basaran04b,Rubio2013} and references therein). Indeed, a recirculation bubble could exist in our case if the negative velocities of the parabolic velocity profile associated with an adverse inner pressure gradient ($\partial P_i/\partial Z>0$), were larger in magnitude than the corresponding positive velocity of the plug flow velocity profile.

In (\ref{Continuidad}), notice that the inner stream pressure $P_i$, which in the slender approximation is constant in the
radial direction, is related to the outer pressure $P_n$ through the normal stress jump
\begin{equation}
P_i=P_n+\sigma \nabla\cdot{\bf n}-
2\mu_o\left[\dfrac{\partial V_n}{\partial R}(R=R_j)-\dfrac{S}{R_j^2}\right]\,,\label{Pi}
\end{equation}
where the term corresponding to the normal viscous stresses associated with the inner stream has been neglected due to the fact that $\lambda\ll 1$. Moreover $\nabla\cdot{\bf n}$ is the interfacial curvature, given by
\begin{equation}
\nabla\cdot{\bf n}=\dfrac{1}{R_j(1+\dot{R}^2_j)^{1/2}}-
\dfrac{\ddot{R}_j}{(1+\dot{R}^2_j)^{3/2}}\,,\label{curvatura}
\end{equation}
with dots denoting derivatives with respect to $Z$. In equation~\eqref{Pi}, notice also that we have taken into account the fact that the velocity field created by the presence of the jet does not contribute to modify the outer pressure field. Indeed, the velocity field $\mathbf{U'}$ generated by a distribution of sources can be expressed in terms of the gradient of a velocity potential, namely, $\mathbf{U'}=\nabla\Phi$ due to the fact that a single three dimensional source and thus, a superposition of sources, are solutions of the Laplace equation. Therefore, since in the zero Reynolds number limit the momentum equation reads
\begin{equation}
-\nabla P'+\mu_{o}\nabla\,^{2}\textbf{U}'=-\nabla
P'-\mu_{o}\nabla\times\left(\nabla\times\textbf{U}'\right)=-\nabla
P'-\mu_{o}\nabla\times\left(\nabla\times\nabla\Phi\right)=0\rightarrow -\nabla
P'=0\, ,
\end{equation}
the only contribution to the outer pressure field comes from the unperturbed flow field, $P_n$. Consequently, the outer axial pressure gradient in the near-axis region, where the
jet surface is located, is calculated by means of the axial projection of the momentum equation,
\begin{equation}
%\begin{split}
-\dfrac{\partial P_n}{\partial Z}=-\mu_o\left[\frac{\partial^2 U_n}{\partial\,Z^2}+\frac{1}{R}\frac{\partial}{\partial\,R}\left(R\frac{\partial\,U_n}{\partial\,R}\right)\right]\rightarrow
-\dfrac{\partial P_n}{\partial Z}=
-\mu_o\left(\dfrac{{\rm d}^2 U_x}{{\rm d}Z^2}+R^2\dfrac{{\rm d}^2 F}{{\rm d}Z^2}+
4F\right)\,,\label{P0}
%\end{split}
\end{equation}
where the expression for $U_n$, given by equation (\ref{Un}), has been used.
Taking $R_i$, $R_i/U_o$ and $\mu_oU_o/R_i$ as characteristic scales for length, time and pressure
respectively, substituting~\eqref{P0} and~\eqref{S} into the
$Z$-derivative of~\eqref{Pi}, and introducing the final expression into~\eqref{Continuidad},
yields the following dimensionless partial differential equation for the jet radius,
\begin{equation}
\begin{split}
&\dfrac{\partial r_j}{\partial t}\left[2r_j+\dfrac{1}{4\lambda}
\left(2r_j\left(\dfrac{\partial r_j}{\partial z}\right)^2+r^2_j\,
\dfrac{\partial^2 r_j}{\partial z^2}\right)\right]-\dfrac{1}{2\lambda}\,r^2_j\,
\dfrac{\partial r_j}{\partial z}\dfrac{\partial}{\partial z}
\left(\dfrac{\partial r_j}{\partial t}\right)-\dfrac{1}{4\lambda}\,r^3_j\,
\dfrac{\partial^2}{\partial z^2}\left(\dfrac{\partial r_j}{\partial t}\right)+\\
&+\dfrac{\partial}{\partial z}\left[\left(u_x+f r_j^2\right)r_j^2+
\dfrac{1}{8\lambda}\left(-3r^4_j\dfrac{{\rm d}^2 u_x}{{\rm d}z^2}-3\,r^6_j\,
\dfrac{{\rm d}^2 f}{{\rm d}z^2}-
4fr^4_j-\Ca^{-1}\,r^4_j\dfrac{\partial\mathcal{C}}{\partial z}-6r^5_j\,
\dfrac{\partial r_j}{\partial z}\dfrac{{\rm d}f}{{\rm d}z}-\right.\right.\\
&\left.\left.-2\left(\dfrac{\partial r_j}{\partial z}\right)^2\,fr^4_j+
2r^2_j\left(\dfrac{\partial r_j}{\partial z}\right)^2\,u_x-2r^3_j
\dfrac{\partial r_j}{\partial z}\dfrac{{\rm d}u_x}{{\rm d}z}-2 u_x r^3_j
\dfrac{\partial^2 r_j}{\partial z^2}-2 f r^5_j
\dfrac{\partial^2 r_j}{\partial z^2}\right)\right]=0\, ,\label{Continuidad1}
\end{split}
\end{equation}
with $u_x$ and $f$ in~\eqref{Continuidad1} given in figure \ref{Uaxis}.
>From now on, lower case variables like those appearing in
equation~\eqref{Continuidad1} indicate the dimensionless version of
their dimensional counterparts and $\mathcal{C}$ indicates the non dimensional expression for
the interfacial curvature given by equation~\eqref{curvatura}. Notice that the equation for
the steady jet shape $r_{j0}(z)$ deduced in~\cite{JFM12}, namely
\begin{eqnarray}
&&q=\left(u_x+f r_{j0}^2\right)r_{j0}^2+\dfrac{1}{8\lambda}\left(-3r^4_{j0}
\dfrac{{\rm d}^2 u_x}{{\rm d}z^2}-3\,r^6_{j0}\,\dfrac{{\rm d}^2 f}{{\rm d} z^2}-
4 f r^4_{j0}-\Ca^{-1}\,r^4_{j0}\dot{\mathcal{C}}_0-\right.\nonumber\\
&&\left.-6 r^5_{j0}\,\dfrac{{\rm d}r_{j0}}{{\rm d}z}
\dfrac{{\rm d}f}{{\rm d}z}-2\left(\dfrac{{\rm d}r_{j0}}{{\rm d}z}\right)^2\,f r^4_{j0}
+2 r^2_{j0}\left(\dfrac{{\rm d}r_{j0}}{{\rm d}z}\right)^2\,u_x-\right.\nonumber\\
&&\left.-2 r^3_{j0}\dfrac{{\rm d}r_{j0}}{{\rm d}z}\dfrac{{\rm d}u_x}{{\rm d}z}-
2 u_x r^3_{j0}\dfrac{{\rm d}^2 r_{j0}}{{\rm d}z^2}-2 f r^5_{j0}
\dfrac{{\rm d}^2 r_{j0}}{{\rm d}z^2}\right)\label{steady}
\end{eqnarray}
can be recovered by simply equating to zero the time derivatives in~\eqref{Continuidad1}.

Let us point out that, in our description, the continuity of tangential stresses across the interface is never invoked since $\lambda\ll 1$. In fact, this condition can be used to calculate the distribution of Stokeslets at the axis, what would introduce a higher order correction ($\sim O[(R_i/L)^2]$) to the velocity field created by the distribution of sources \citep{Buckmaster}. It is also interesting to note that, in our approach to deduce equation (\ref{Continuidad1}), terms of the order of $\sim O[(R_i/L)^2]$ have been neglected in the expression for the outer velocity field given by (\ref{U1}). However, the azimuthal curvature term $\sim \partial^2 R_j/\partial\,Z^2$, which is also of the order of $\sim O[(R_i/L)^2]$, has been retained in equation (\ref{curvatura}). This apparent inconsistency is justified since, as pointed out in \cite{GyC,Basaran1D2D,basaran04b,EggersVillermaux,Rubio2013} and references therein, to improve the agreement with experiments in one dimensional models such as the one 
developed here, the exact expression of the interfacial curvature needs to be retained in the analysis of capillary dominated flows. Also, it can be appreciated in equation (\ref{P0}) that, in order for the contribution of $P_n$ to the inner pressure gradient to possess the same degree of accuracy as the contribution of the capillary pressure, the terms $\partial^2 U_n/\partial Z^2$ and $R^2 \partial^2 F/\partial Z^2$, which are $\sim O[(R_i/L)^2]$ when compared to $F$, have also been retained in the analysis.

\subsection{Global stability analysis}
\label{subsec:gla}

To find whether the solutions of equation~\eqref{steady} are stable or not, we substitute
the ansatz
\begin{equation}
r_j=r_{j0}(z)+r_{j1}(z,t)\label{rj0rj1}\,,
\end{equation}
into equation~\eqref{Continuidad1} and retain only linear terms in $r_{j1}(z,t)$ and its
derivatives, yielding the following linear partial differential equation for the perturbed
jet radius,
\begin{equation}
\begin{split}
&\dfrac{\partial r_{j1}}{\partial t}\left[2 r_{j0}+\dfrac{1}{4\lambda}
\left(2 r_{j0}\dot{r}_{j0}^2+r^2_{j0} \ddot{r}_{j0}\right)\right]-
\dfrac{1}{2\lambda}\,r^2_{j0}\dot{r}_{j0}\dfrac{\partial}{\partial z}
\left(\dfrac{\partial r_{j1}}{\partial t}\right)-\dfrac{1}{4\lambda}\,r^3_{j0}\,
\dfrac{\partial^2}{\partial z^2}\left(\dfrac{\partial r_{j1}}{\partial t}\right)+\\
&+r_{j1}B_0+\dfrac{\partial r_{j1}}{\partial z}B_1+
\dfrac{\partial^2 r_{j1}}{\partial z^2}B_2+\dfrac{\partial^3 r_{j1}}{\partial z^3}B_3+
\dfrac{\partial^4 r_{j1}}{\partial z^4}B_4=0\,,\label{Continuidadresumida}
\end{split}
\end{equation}
where the $B_j$ are five functions of $r_{j0}(z),f(z),u_x(z)$ and their $z$-derivatives,
whose expressions~\eqref{AB0}-\eqref{AB4} are provided in Appendix~A.

Equation~\eqref{Continuidadresumida} needs to be solved subjected to the boundary condition
$r_{j1}(z=0)=0$, and to the additional boundary condition expressing the fact that the
flow rate at $z=0$ is independent of time and equal to $q$, namely
\begin{equation}
\begin{split}
&\dfrac{\partial r_{j1}}{\partial z}\left[-\dfrac{3}{4\lambda}\,r^5_{j0}\dot{f}-
\dfrac{r^3_{j0}}{4\lambda}\dot{u}_x+\dot{r}_{j0}\left(\dfrac{1}{2\lambda}r_{j0}^2\,u_x-
\dfrac{1}{2\lambda}r^4_{j0}\,f\right)-\dfrac{1}{8\lambda \Ca}
\left(r^2_{j0}\dot{r}_{j0}^2d_0^{-3/2}-\right.\right.\\
&\left.\left.-r^2_{j0}d_0^{-1/2}+3r^3_{j0}\dot{r}^2_{j0}\ddot{r}_{j0}\,d_0^{-5/2}-r^3_{j0}
\ddot{r}_{j0}\,d_0^{-3/2}-15\dot{r}_{j0}^2\ddot{r}_{j0}^2\,r^4_{j0}\,d_0^{-7/2}+\right.\right.\\
&\left.\left.+3\ddot{r}_{j0}^2\,r_{j0}^4\,d_0^{-5/2}+3\,r^4_{j0}\dot{r}_{j0}\,r^{(3)}_{j0}\,
d_0^{-5/2}\right)\right]+
\dfrac{\partial^2 r_{j1}}{\partial z^2}\left[-\frac{u_x}{4\lambda}\,r^3_{j0}-\right.\\
&\left.-\dfrac{f}{4\lambda}r^5_{j0}-\dfrac{1}{8\lambda \Ca}\left(-r^3_{j0}\dot{r}_{j0}\,
d_0^{-3/2}+6r^4_{j0}\dot{r}_{j0}\ddot{r}_{j0}\,d_0^{-5/2}\right)\right]+
\dfrac{\partial^3 r_{j1}}{\partial z^3}\,\dfrac{1}{8\lambda \Ca}\,r_{j0}^4\,d_0^{-3/2}=0\,.
\label{CC1}
\end{split}
\end{equation}
Equation~\eqref{CC1} has been deduced retaining only linear terms in $r_{j1}$ and its
derivatives in the expression resulting from the substitution of the ansatz~\eqref{rj0rj1}
into equation~\eqref{steady}.

Writing the perturbed radius as
\begin{equation}
r_{j1}(z,t)=e^{\omega t}\,\bar{r}_{j1}(z)\,,\label{Omega}
\end{equation}
the different eigenvalues $\omega$ and their corresponding global eigenfunctions
$\bar{r}_{j1}(z)$ are found numerically through a Chebychev spectral collocation method applied
to the expression resulting from the substitution of the ansatz~\eqref{Omega}
into~\eqref{Continuidadresumida}. For that purpose, the physical space $0\leq z\leq l_{\infty}$,
with $l_{\infty}\gg 1$ the dimensionless jet length, is mapped into the Chebychev space
$-1\leq \xi\leq 1$ through the transformation
\begin{equation}
z=\dfrac{c\,(1-\xi)}{1-\xi^2+\frac{2c}{l_\infty}}\,.\label{Transformacion}
\end{equation}
To cluster as many points as possible in the cone-jet transition region, which is where the
interfacial curvature suffers the strongest variations, the constant $c$ in~\eqref{Transformacion}
is chosen to be twice the distance from the injection nozzle where the curvature of the steady
jet shape is minimum. Forcing the discrete version of equation~\eqref{Continuidadresumida} to be
satisfied at the $N-2$ Chebychev collocation points within the physical domain (all the
collocation points except those at the boundaries, namely $z=0$ and $z=l_\infty$) and imposing
at $z=0$ the boundary condition $\bar{r}_{j1}(z=0)=0$ as well as the one given by the discrete
version of the equation resulting from the substitution of~\eqref{Omega} into~\eqref{CC1},
yields the following linear system of equations for the $N$ eigenvalues and their corresponding
eigenfunctions,
\begin{equation}
\omega\mathcal{A}\cdot\mathbf{\bar{r}_{j1}}=\mathcal{B}\cdot\mathbf{\bar{r}_{j1}}\,,\label{Eceig}
\end{equation}
where the eigenvector $\mathbf{\bar{r}_{j1}}$ contains the values of the perturbed jet radius at
the $N$ collocation points. The $N\times N$ matrices $\mathcal{A},\mathcal{B}$ in
equation~\eqref{Eceig} are respectively given by
\begin{equation}
\mathcal{A}=\mathrm{diag}\left(-\frac{1}{4\lambda}\,r^3_{j0}\right)\mathcal{D}^2-
\mathrm{diag}\left(\frac{1}{2\lambda}\,r^2_{j0}\dot{r}_{j0}\right)\mathcal{D}+
\mathrm{diag}\left(2\,r_{j0}+\frac{1}{4\lambda}\left(2\,r_{j0}\dot{r}_{j0}^2+r^2_{j0}\,
\ddot{r}_{j0}\right)\right)\mathcal{I}\,,\label{A}
\end{equation}
and
\begin{equation}
\mathcal{B}=\mathrm{diag}\left(B_4\right)\mathcal{D}^4+\mathrm{diag}\left(B_3\right)\mathcal{D}^3+
\mathrm{diag}\left(B_2\right)\mathcal{D}^2+\mathrm{diag}\left(B_1\right)\mathcal{D}+
\mathrm{diag}\left(B_0\right)\,,
\end{equation}
where $\mathrm{diag}(\cdot)$ indicates the diagonal matrix whose elements are the values of the
$z$-dependent argument particularized at the $N$ collocation points, $\mathcal{I}$ indicates the
identity matrix, and the column vector
\begin{equation}
\mathcal{D}^n\cdot \mathbf{\bar{r}_{j1}}
\end{equation}
contains the value of the $n$-th derivative of $\bar{r}_{j1}$ in the physical space at the $N$
collocation points. Let us point out that our stability results were independent of
the numerical jet length, $l_{\infty}$, provided that $l_{\infty}\gg 1$.

\begin{figure}
\centering
\includegraphics[width=1\textwidth]{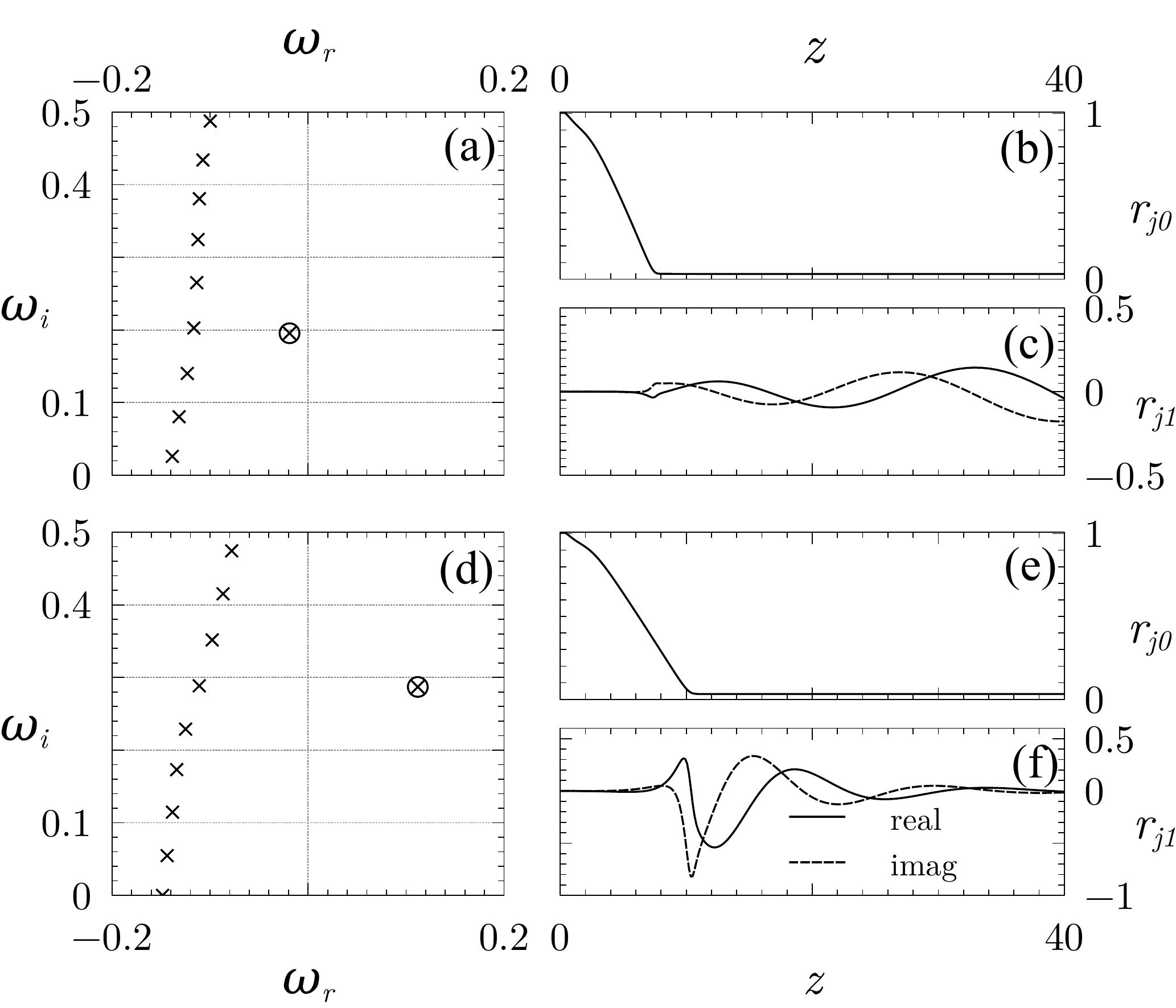}%
\caption{(a,d) Spectra, (b,e) unperturbed jet radius and (c,f) eigenfunctions associated with
the oscillatory global mode 1 for $\lambda=0.01$, $q=10^{-3}$ and $\Ca=1>\Ca^*$ (globally
stable flow, a--c), and $\Ca=0.8<\Ca^*$ (globally unstable flow, d--f). From
figure~\ref{fig4}d, notice that $\omega_i>0$ when the real part of the leading eigenvalue is
$\omega_r>0$. Also, from figure~\ref{fig4}f, notice that the associated eigenfunction
experiences spatial oscillations. Thus, the unstable global mode of type 1 represents a
perturbation that grows and propagates in both space and time. Crosses indicate the position of the eigenvalues, whereas the leading eigenvalue (the one corresponding to the largest growth rate), is encircled. \label{fig4}}
\end{figure}

\begin{figure}
\centering
\includegraphics[width=1\textwidth]{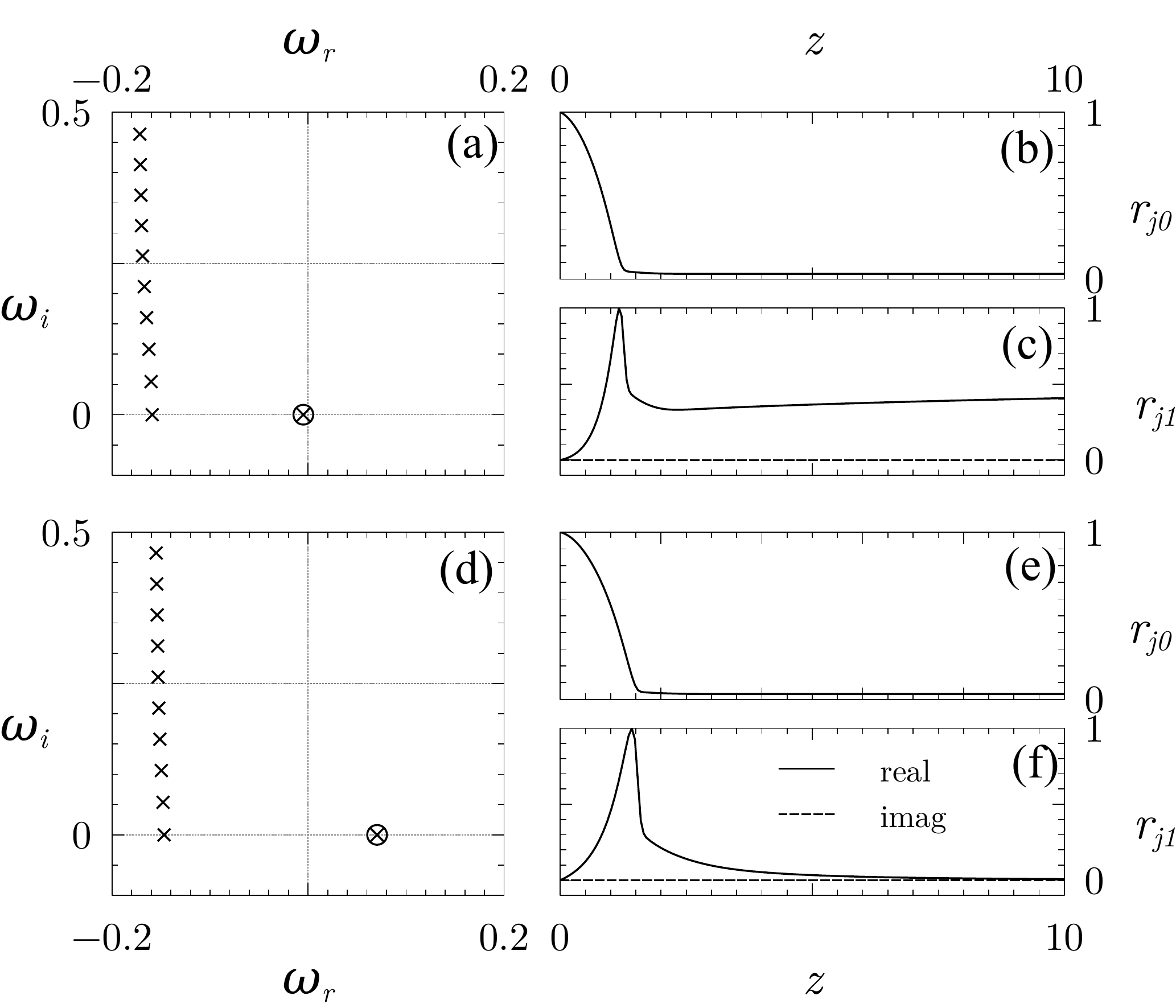}%
\caption{(a,d) Spectra, (b,e) unperturbed jet shape and (c,f) eigenfunctions associated with
the non-oscillatory global mode 2 for $\lambda=0.1$, $q=10^{-3}$ and values of $\Ca=1>\Ca^*$
(stable flow, a--c), and $\Ca=0.8<\Ca^*$ (unstable flow, d--f). From figure~\ref{fig5}d,
notice that $\omega_i=0$ when the real part of the leading eigenvalue is $\omega_r>0$.
Also, from figure~\ref{fig5}f, notice that the associated eigenfunction does not experience
any kind of spatial oscillations. Thus, the unstable global mode of type 2 represents the
temporal growth of a perturbation which does not oscillate neither in time nor in space. Crosses indicate the position of the eigenvalues, whereas the leading eigenvalue (the one corresponding to the largest growth rate), is encircled.
\label{fig5}}
\end{figure}

The steady solution $r_{j0}(z)$, which is found solving the nonlinear equation~\eqref{steady}
by means of a Newton-Raphson method, will be globally stable when the real parts of \emph{all}
the eigenvalues are negative, whereas it will be globally unstable if the real part of at least
one eigenvalue, is positive. Thus, to find the function $\Ca^*(\lambda,q)$ we proceed as follows:
for given values of $\lambda$ and $q$, we first find the unperturbed jet shape $r_{j0}(z)$ by
solving equation~\eqref{steady} for a sufficiently large value of $\Ca$. The good agreement of
the resulting function $r_{j0}(z)$ with experiments shown in~\citet{JFM12}, constitutes the
first evidence that supports the accuracy of our generalized slender body theory. Once the
steady solution is found, the matrices $\mathcal{A},\mathcal{B}$ in equation~\eqref{Eceig} can
be computed since they only depend on $r_{j0}$, its derivatives and on the control parameters
$(\lambda,\Ca,q)$. Then, we proceed to find the $N$ eigenvalues and their corresponding
eigenvectors using standard Matlab functions (see figures~\ref{fig4} and~\ref{fig5}). If the
real parts of all eigenvalues are negative, as in the cases shown in figures~\ref{fig4}a
and~\ref{fig5}a, $\Ca$ is decreased and the full process is repeated again until the real part
of one of the eigenvalues crosses zero, as illustrated in figures~\ref{fig4}d and~\ref{fig5}d.
Note that, for given values of $\lambda$ and $q$, the value of $\Ca^*$ is determined by the
condition that the real part of the leading eigenvalue is zero.

\begin{figure}
\centering
\includegraphics[width=1\textwidth]{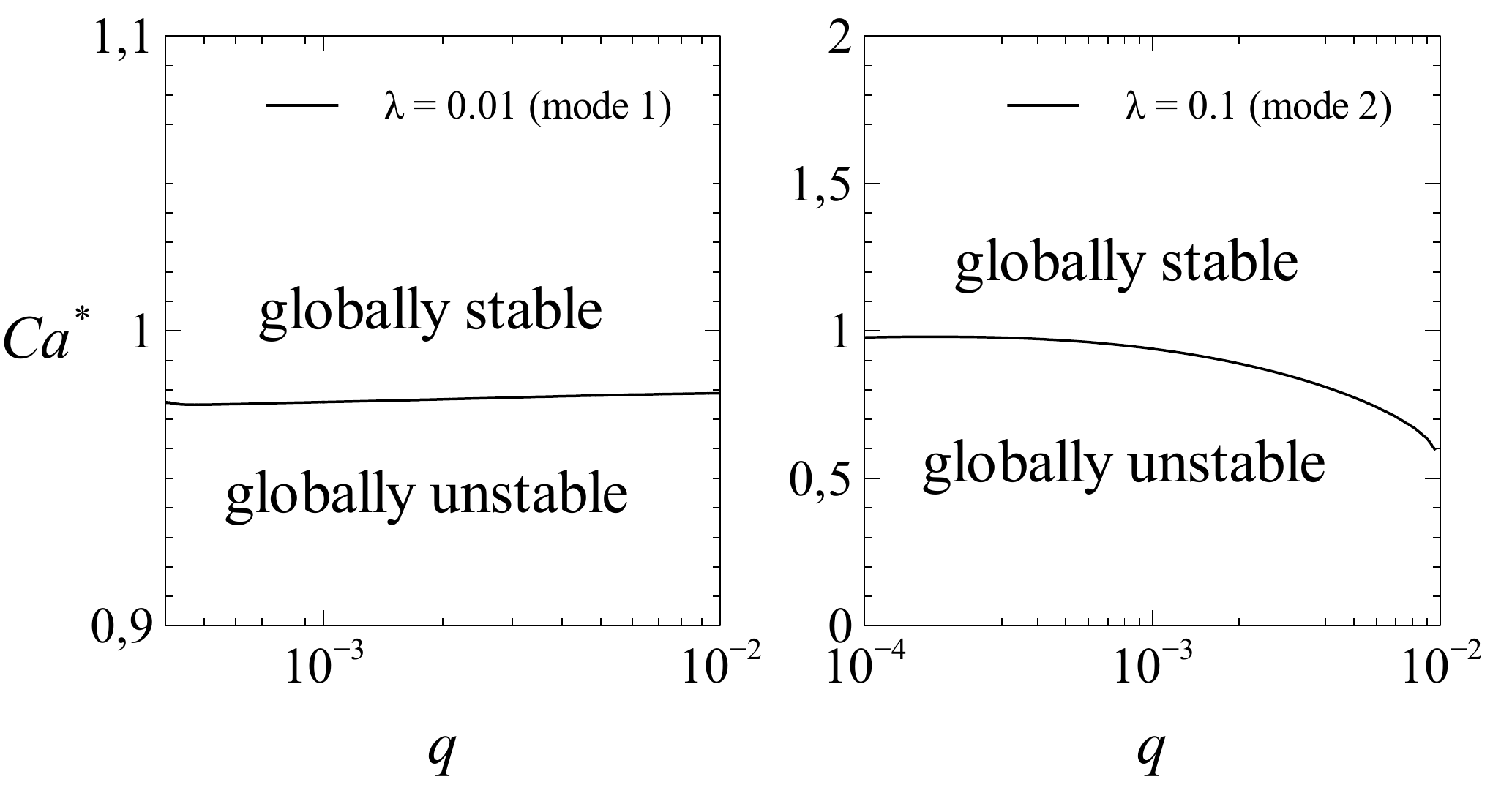}%
\caption{Dependence of $\Ca^*$ on $q$ for the oscillatory mode 1 (left plot), and the
non-oscillatory mode 2 (right plot).\label{fig6}}
\end{figure}

\begin{figure}
\centering
\includegraphics[width=1\textwidth]{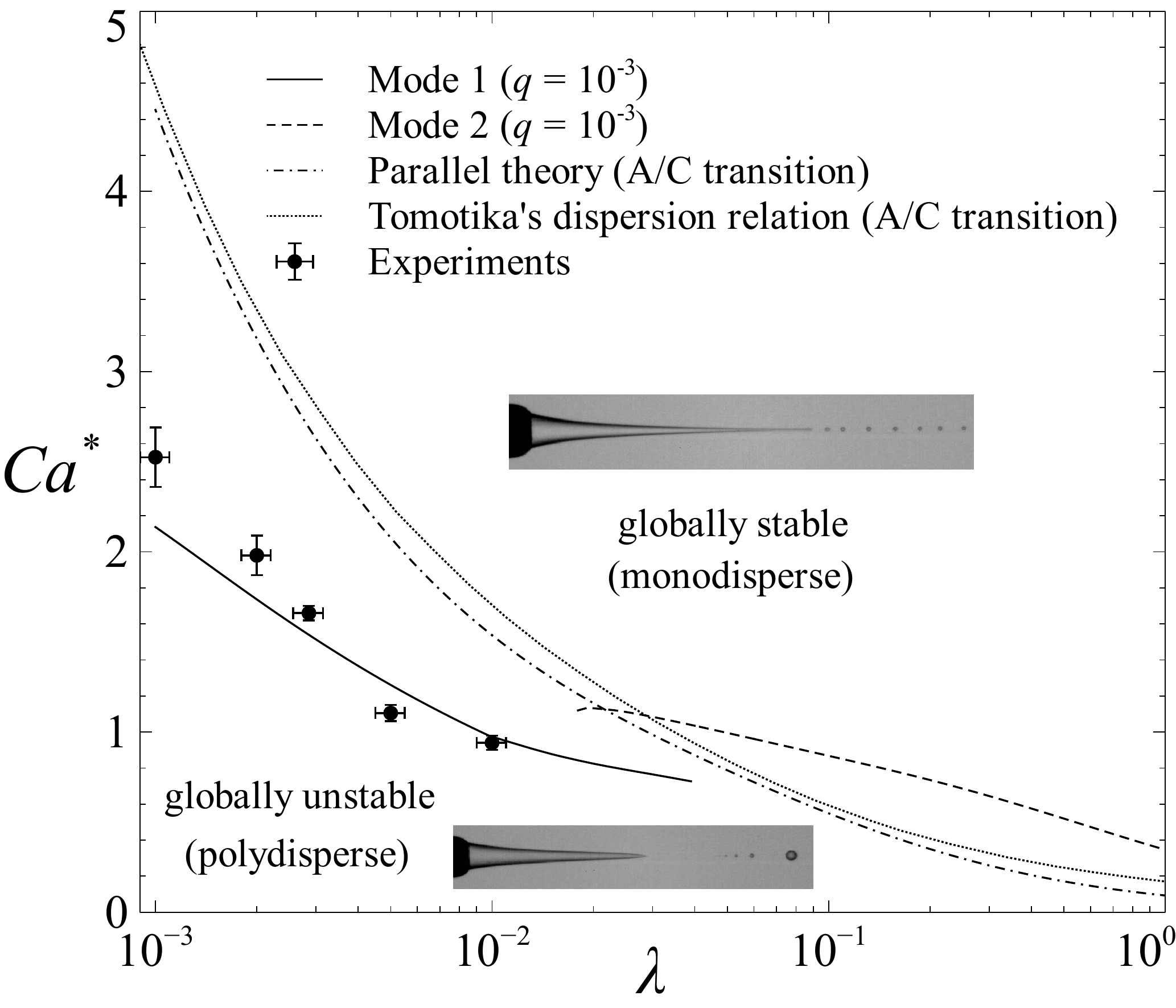}%
\caption{The critical capillary number, $\Ca^*$, as a function of the viscosity ratio
$\lambda$. Symbols show the experimental transition points separating the monodisperse
($\Ca>\Ca^*$, figures~\ref{fig1}a-e), and polydisperse ($\Ca<\Ca^*$, figures~\ref{fig1}f-j)
droplet generation processes. The solid and dashed lines represent the instability thresholds
of the global modes 1 (oscillatory) and 2 (non-oscillatory), respectively. The dash-dotted
and dotted lines correspond to the absolute/convective (A/C) transition, computed applying
the zero group velocity condition to equation~\eqref{reldisp} in the limit $z\to\infty$, and to
Tomotika's dispersion relation, respectively.\label{fig7}}
\end{figure}

We have identified two different types of eigenfunctions that represent either a spatio-temporally
oscillating mode (mode 1, see figure~\ref{fig4}) or a non-oscillatory mode (mode 2, see figure~\ref{fig5}). Both
families are also differentiated by the fact that, contrarily to the case of the non-oscillatory
mode, the amplitude of the oscillatory eigenfunction is virtually zero in the cone region, as it
can be deduced from figure~\ref{fig4}. At this point, let us anticipate that, within the ranges of
$\lambda$ and $q$ explored in this study, the stability analysis predicts that the oscillatory
mode dominates over the steady one, so it is reasonable to expect that the shape of the
eigenfunction depicted in figure~\ref{fig4}f is somehow related with our experimental
observations. This is indeed the case since, as revealed by figures~\ref{fig1} and~\ref{fig2},
the cone region is unperturbed even in the cases in which the drop formation process is not
periodic (see also the movies provided as supplementary material). Moreover, the maximum amplitude of the eigenfunction
associated with the oscillatory mode is located slightly downstream the cone tip, and this is
consistent with the experimental observation that the aperiodic formation process is associated
with an unsteady emission of drops from a region located very close to the cone apex. Our global
stability analysis also reveals that the value of the critical capillary number is hardly dependent
on $q$ for values of the viscosity ratio $\lambda\lesssim 0.1$ (see figure~\ref{fig6}). This
finding is also in agreement with the experimental evidence shown in figure~\ref{fig2}, where it
can be appreciated that $\Ca^*$ is rather insensitive to changes in $q$. Therefore, in view of the
slight variation of $\Ca^*$ with $q$ shown in figure~\ref{fig6}, the dependence of the critical
capillary number with $\lambda$ is calculated for a fixed value of $q=10^{-3}$. The result of this
calculation, depicted in figure~\ref{fig7}, compares favorably with experiments, a fact that
further supports our theory. Since our theoretical predictions are much closer to the experimental
observations than those obtained assuming that the jet is cylindrical, we can conclude that the
strong variation of the jet radius in the axial direction plays an essential role in stabilizing
the drop formation process.

We would like to emphasize that our analysis is able to predict the transition from the regular
drop formation process that takes place when a thin steady ligament is issued from the cone apex
(jetting regime) to the regime in which droplets detach aperiodically at the cone-jet region
(dripping regime), but not the reverse dripping to jetting transition. Note also that the
destabilization of the regular drop formation regime observed in our experiments shares some similarities with
the transition from jetting to dripping observed in liquid jets stretched by
gravity~\citep{clanet99,Basaran1D2D,basaran04b}, where the dripping regime manifests itself by the formation of drops right from or slightly downstream the exit of the injection tube. The main difference with the gravitational case is that, in the physical situation analyzed here, dripping occurs right from the tip of the conical drop, whose global shape is stable except very close to the microdrop emission region. The local oscillations appearing at the tip explain why the emission process is not periodic and leads to the formation of unevenly sized drops when the flow is globally unstable (see figures 1f-j, 6f and the movies provided as supplementary material). Let us point out that we also observe the formation of drops right from the exit tube, but this regime is found for values of the capillary number notably smaller than those investigated here \citep{SuryoBasaran}. This type of dripping regime, which is much more similar to that observed in the gravitational case, has not been reported here since the 
diameters of the drops obtained are imposed by the geometry of the injector and, thus, are not of interest for applications due to the fact that their sizes are of the order of 100 microns or even larger.

\subsection{Comparison with Tomotika's stability analysis and elucidation of the physical mechanism leading to the higher stability exhibited by stretched jets}

There is still a remaining question that needs to be answered: why do our stretched microjets
exhibit a more stable behavior than their cylindrical counterparts? At first sight it could
be thought that, since ${\rm d}u_x/{\rm d}z>0$~ (see figure \ref{Uaxis}), the growth
rate of perturbations decreases thanks to the kinematic mechanism described in~\citet{Tomotika2}
and in~\citet{FrankelWeihs1985}, whereby the amplitude of the perturbations decreases while
their wavelenght increases due to the fact that fluid particles are elongated under the action
of the outer straining flow field. However, we will show below that, although the jet is indeed
stabilized thanks to this kinematic mechanism, the main reason for the higher stability exhibited
by the stretched microjets of figures~\ref{fig1} and~\ref{fig2} is in fact related with the
gradient of capillary pressure.

Indeed, let us adopt the slender approximation of assuming that the wavelength of the perturbation,
$2\pi/k$, with $k$ the wavenumber, is much smaller than the length along which $r_{j0}$ experiences
variations of its same order of magnitude. This condition, clearly verified for $z\gg 1$, permits
to look for traveling-wave solutions to equation~\eqref{Continuidadresumida}, of the form
$r_{j1}=e^{i\left(kz-\omega\,t\right)}\bar{\bar{r}}_{j1}$, with $k\in \mathcal{R}$ and
$\bar{\bar{r}}_{j1}$ slowly varying functions of $z$. At leading order, this approach provides the
following dispersion relation for $\omega$ as a function of $k$,
\begin{equation}
 \begin{split}
&D(\omega,k)=-i\omega\left(2 r_{j0}+\dfrac{r^3_{j0}}{4\lambda}\,k^2\right)+
\left(\dfrac{b_{c0}}{\Ca}+2 r_{j0}\dot{u}_x\right)+\\
&+\left[2 r_{j0} u_x+r^3_{j0}\left(f\left(4-\dfrac{2}{\lambda}\right)-
\dfrac{7}{4\lambda}\ddot{u}_x\right)\right]ik+\dfrac{r^2_{j0}}{8\lambda\Ca}\,k^2+\\
&+\left(\dfrac{r^3_{j0}}{2 \lambda}\,u_x+\dfrac{r^5_{j0}}{4\lambda}\,f\right)ik^3+
\dfrac{r^4_{j0}}{8\lambda\Ca}\,k^4=0\,,\label{reldisp}
\end{split}
\end{equation}
where the term $b_{c0}\in \mathcal{R}$, provided in equation~\eqref{bc0} of Appendix~A, arises
when linear perturbations in the expression for the \emph{full curvature} of the jet are retained.
Neglecting all the $z$-derivatives of $r_{j0}$, what would correspond to the case of a purely
cylindrical jet or to the solution of~\eqref{steady} at $z\rightarrow \infty$, leads to $u_x=1$
and to $b_{c0}=\dot{u}_x=\ddot{u}_x=f=0$. The resulting dispersion relation ~\eqref{reldisp} in the limit $z\rightarrow \infty$ is identical to Tomotika's long wave limit dispersion relation deduced by ~\cite{PRLGoldstein,PoFPowers}, a fact constituting a further proof of our theory. Applying the zero group velocity condition
$\partial\omega/\partial k=0$, $\omega_i=0$ to equation~\eqref{reldisp} in the limit $z\rightarrow \infty$,
the boundary separating the absolutely and convectively unstable states can be calculated as
explained in~\cite{AnnRevHuerre,Gordillo01c,PoF05}. The resulting A/C transition curve,
shown in figure~\ref{fig7} (dash-dotted line), is very close to that obtained from the much more
involved calculation~\citep{PRLGoldstein,PoFPowers,PREInnombrable} that makes use of Tomotika's dispersion
relation (dotted line).

\begin{figure}
\centering
\includegraphics[width=0.8\textwidth]{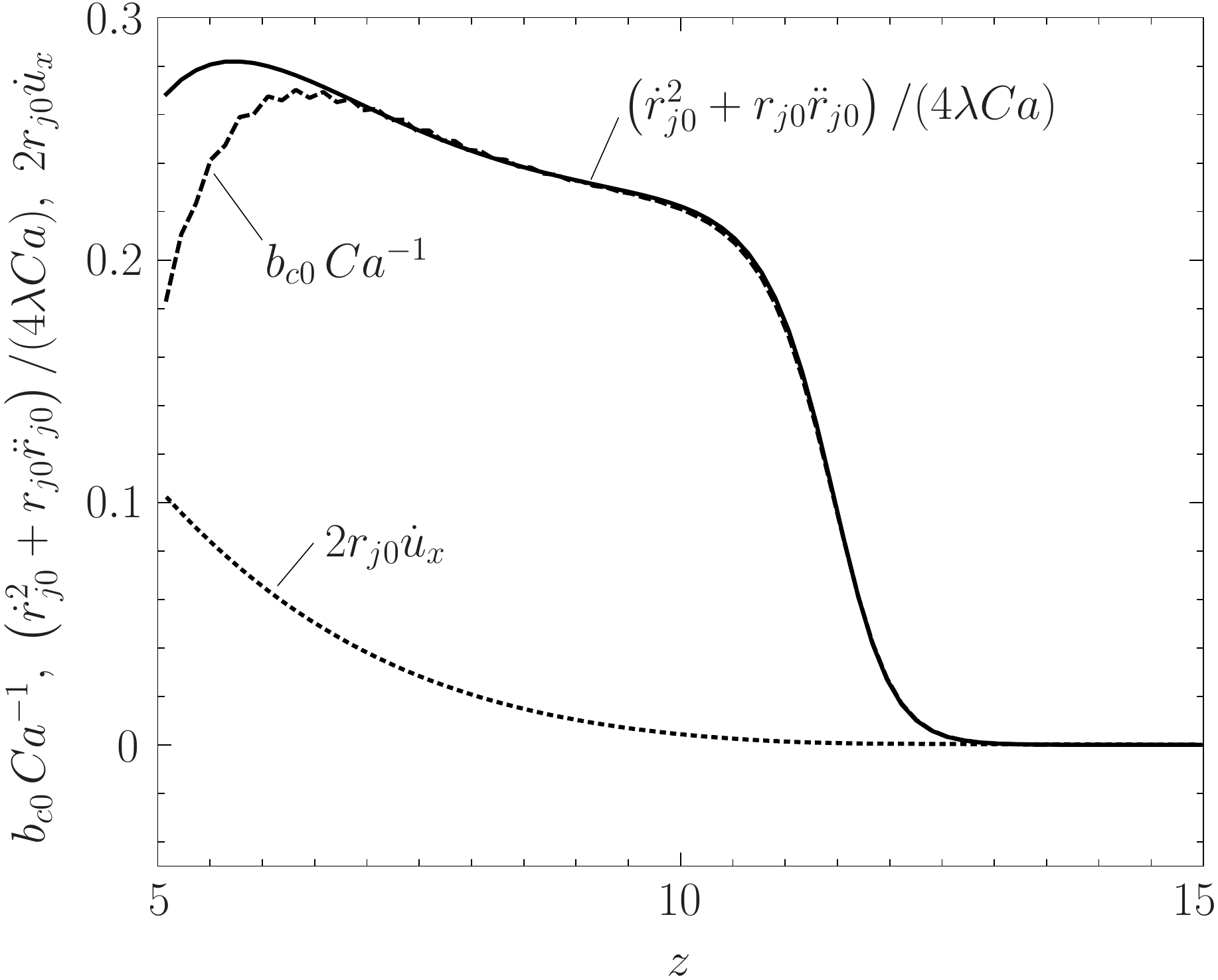}%
\caption{This figure shows, for $\lambda=0.01$ and $Ca=1$ that the function $b_{c0}\Ca^{-1}$
in~\eqref{reldisp} (dashed line) can be well approximated as
$(\dot{r}_{j0}^2+r_{j0}\ddot{r}_{j0})/(4\lambda \Ca)$ (solid line), and that $2r_{j0}\dot{u}_x$
(dotted line), is smaller that $b_{c0}\Ca^{-1}$, meaning that the stabilization mechanism
associated with the action of capillary forces on highly curved interfaces, dominates over the
kinematic stabilization produced by the elongation of fluid
particles~\citep{Tomotika2,FrankelWeihs1985}.
\label{fig8}}
\end{figure}

However, as stated above, to understand the higher stability of stretched jets we must retain
the \emph{non-parallel} terms in the dispersion relation~\eqref{reldisp} associated with axial
variations of $r_{j0}$ and $u_x$.
% let us point out that the larger is the imaginary part of $-i\omega$, i.e., the larger is the
% modulus of the terms proportional to $ik$ and $ik^3$ in~(\ref{reldisp}), the larger is the speed at
% which perturbations are convected downstream. In particular, $f(4-2/\lambda)-7/(4\lambda)\ddot{u}_x>0$,
% a fact indicating that favorable pressure gradients contribute to the downstream convection of
% perturbations. But
The main effect of non-parallel terms is to decrease the real part of $-i\omega$, namely, the
growth rate of perturbations. Indeed, $b_{c0}\Ca^{-1}+2\,r_{j0}\dot{u}_x>0$ since $\dot{u}_x>0$
(see figure \ref{Uaxis}) and $b_{c0}\Ca^{-1}>0$ (see figure~\ref{fig8}). It was already
pointed out above that positive axial velocity gradients stabilize the
jet~\citep{Tomotika2,FrankelWeihs1985,EggersVillermaux}, but figure~\ref{fig8} reveals that
$2r_{j0}\dot{u}_x<b_{c0}\Ca^{-1}$ near the tip of the conical drop, whose steady shape is depicted in figure 6b. The reason why the non-parallel
term $b_{c0}\Ca^{-1}$ contributes to the stabilization of the stretched jet can be easily explained
as follows: the gradients of capillary pressure in~\eqref{Continuidad} produce an axial variation
of the flow rate per unit length given by
$\Delta\,Q_{i,cap}=\sigma/(8\mu_i)\partial\left(R_j^4\,\partial(-\nabla\cdot{\bf n})/
\partial Z\right)/\partial Z$. Using the slender approximation for the curvature, namely,
$\nabla\cdot{\bf n}\sim 1/R_{j}$, linearizing the resulting expression around $R_{j0}$, and
using dimensionless variables, yields
$\Delta q_{i,cap}\sim(\dot{r}_{j0}^2+r_{j0}\ddot{r}_{j0})\,r_{j1}/(4\lambda \Ca)$ which, as can
be appreciated in figure~\ref{fig8}, is an excellent approximation to $b_{c0}\Ca^{-1}$ in the
cone-jet transition region. Therefore, since $\dot{r}_{j0}^2+r_{j0}\ddot{r}_{j0}>0$, perturbations
such that $r_{j1}>0$ (resp. $r_{j1}<0$), that tend to increase (resp. decrease) the jet radius,
cause $\Delta\,q_{i,cap}>0$ (resp. $\Delta\,q_{i,cap}<0$), leading to
$\partial r_{j1}/\partial t<0$ (resp. $\partial r_{j1}/\partial t>0$) by virtue of the mass
balance~\eqref{Continuidad}, thus explaining the stabilizing effect of the gradient of capillary
pressure existing inside stretched jets.

\section{Conclusions}
\label{sec:conclusions}

In this paper we study the effect of the capillary number on the size distribution of the
drops composing the microemulsions obtained when two immiscible liquid streams with $\lambda\ll 1$ co-flow at
low Reynolds numbers and highly stretched liquid threads are
produced~\citep{SuryoBasaran,Marin}. Our experiments reveal that, when $\Ca\gtrsim O(1)$
and the flow rate ratio is such that $q\ll 1$, a liquid jet with a diameter substantially
smaller than that of the injection tube is issued from the tip of a conical drop pinned
at the exit of the injector. We find that this ligament is unsteady and breaks into unevenly
sized droplets for $\Ca<\Ca^*(\lambda)$ with $\Ca^*\sim O(1)$ the critical capillary number
and that, thanks to the fact that the growing capillary perturbations are convected downstream
the cone-jet transition region, the thin liquid jet is steady and breaks into uniformly
sized drops when $\Ca>\Ca^*(\lambda)$. With the purpose of predicting the dependence of the
critical capillary number on $\lambda$, we have first determined the values of $\Ca$ for which the group velocity of neutral waves is zero under the parallel flow assumption. The
results of this analysis using the dispersion relation which describes the growth of
capillary perturbations in cylindrical capillary jets~\citep{Tomotika}, reveal that the
boundary separating the absolutely unstable region from the convectively unstable
one~\citep{PRLGoldstein,PoFPowers,PREInnombrable} is well above the experimental values of the
critical capillary number, a fact indicating that the non-cylindrical liquid ligaments of our experiments are more
stable than their cylindrical counterparts. Thus, to improve the agreement between
experiments and theory, we have performed a global stability analysis retaining the highly
stretched shape of the liquid jet as well as realistic inner and outer velocity fields.
For that purpose, following the slender body theory developed in~\citet{JFM12}, we have
deduced the partial differential equation~\eqref{Continuidad1} that describes the
spatiotemporal evolution of the jet radius and whose steady limit accurately reproduces
the shape of the unperturbed liquid ligaments. Also, we have shown here that our equation~\eqref{Continuidad1} is able to reproduce Tomotika's stability analysis in the limit of cylindrical jets of a low viscosity fluid, a fact that further supports the validity of our theory. Moreover, the results of our stability analysis, which
are in fair agreement with experimental measurements, indicate that the regime in which
monodisperse emulsions are produced corresponds to those values of the capillary number for which
the cone-jet system is \emph{globally stable}. We can thus conclude that the
gradients of interfacial curvature and of axial velocity play an essential role in the
propagation of perturbations along stretched capillary jets. Notice that this idea has also been successfully applied to describe the jetting to dripping transition experienced by gravitationally stretched capillary jets using the one-dimensional equations for the jet radius and the axial velocity~\citep{Rubio2013}. Let us point out that the simplicity of the one dimensional approximation used here to study the stability of stretched jets contrasts with the difficulty of analyzing the stability of two dimensional or three dimensional free surface flows \citep[see e.g.][]{Christodoulou}.

One of our major findings here is the discovery of a new stabilization mechanism that
resorts to the existence of capillary pressure gradients inside low Reynolds number
stretched jets that differs from the one based on the elongation of fluid
particles~\citep{Tomotika2}. To conclude, we would like to emphasize that equation~\eqref{Continuidad1}
is rather general in the sense that it can be applied to describe the propagation and
growth of capillary perturbations along the jets produced, for instance,
in flow-focusing devices operated at low Reynolds numbers. Indeed, notice that the influence of each
particular geometry in the equation for the jet radius~\eqref{Continuidad1} comes through
the functions $u_x$ and $f$, which can be easily found numerically by solving Stokes
equations subjected to the appropriate boundary conditions when the inner fluid is removed from the flow domain. Currently, we are extending the idea of decomposing the flow field as the addition of two simpler velocity fields to those cases in which the Reynolds number of the outer stream is large. The same type of ideas based on the slender body approach could also be applied to describe the stability of electrohydrodynamic tip streaming regimes that, under appropriate conditions, take place when electrical stresses act on the interface of liquid threads with a cone-jet structure \citep{DelaMora,BarreroAnnurev,Collins1,Collins2}.

\begin{acknowledgements}
JMG and FCC thank financial support by the Spanish MINECO under Project DPI2011-28356-C03-01
and the Junta de Andaluc\'ia under Project P08-TEP-03997. AS thanks financial support by the
Spanish MINECO under Project DPI2011-28356-C03-02. These research projects have been partly
financed through European funds. The authors wish to express their gratitude to Elena de
Castro-Hen\'andez for providing them with the numerical values of functions $f$ and $u_x$
and for her careful revision of the equations in the text.
\end{acknowledgements}

\appendix
\section{}

The partial differential equation~(\ref{Continuidadresumida}) for the perturbed jet radius
$r_{j1}(z,t)$ is the compact form of the following expression:
\begin{equation}
\begin{split}
&\dfrac{\partial r_{j1}}{\partial t}\left[2 r_{j0}+\dfrac{1}{4\lambda}\left(2 r_{j0}\dot{r}_{j0}^2+
r^2_{j0}\,\ddot{r}_{j0}\right)\right]-\dfrac{1}{2\lambda}\,r^2_{j0}\dot{r}_{j0}
\dfrac{\partial}{\partial z}\left(\dfrac{\partial r_{j1}}{\partial t}\right)-
\dfrac{1}{4\lambda}\,r^3_{j0}\,\dfrac{\partial^2}{\partial z^2}\left(
\dfrac{\partial r_{j1}}{\partial t}\right)+\\
&+r_{j1}\left[\dfrac{b_{c0}}{\Ca}+2 r_{j0}\dot{u}_x+4r^3_{j0}\dot{f}-r^3_{j0}
\left(\dfrac{3}{2\lambda}u^{(3)}_x+\dfrac{2}{\lambda}\dot{f}\right)-r^5_{j0}
\dfrac{9}{4\lambda}f^{(3)}+\right.\\
&+\left.\dot{r}_{j0}\left(2\,u_x-\dfrac{21}{4\lambda}r^2_{j0}\ddot{u}_x+fr^2_{j0}
\left(12-\dfrac{6}{\lambda}\right)-\dfrac{15}{\lambda}r^4_{j0}\ddot{f}\right)+
\dot{r}_{j0}^2\left(-\dfrac{16}{\lambda}r^3_{j0}\dot{f}-\dfrac{1}{\lambda}
r_{j0}\dot{u}_x\right)\right.+\\
&+\left.\dot{r}_{j0}^3\left(-\dfrac{3}{\lambda}r^2_{j0}\,f+\dfrac{1}{2\lambda}u_x\right)
-\ddot{r}_{j0}\left(\dfrac{5}{\lambda}r^4_{j0}\dot{f}+\dfrac{3}{2\lambda}r^2_{j0}
\dot{u}_x\right)-\dot{r}_{j0}\ddot{r}_{j0}\left(\dfrac{7}{\lambda}r^3_{j0}\,f+
\dfrac{1}{2\lambda}u_x\,r_{j0}\right)-\right.\\
&-\left.r^{(3)}_{j0}\left(\dfrac{3}{4\lambda}u_xr^2_{j0}+\dfrac{5}{4\lambda}
r^4_{j0}\,f\right)\right]+\dfrac{\partial r_{j1}}{\partial z}\left[\dfrac{b_{c1}}{\Ca}+
2r_{j0}u_x+r^{3}_{j0}\left(f\left(4-\dfrac{2}{\lambda}\right)-\dfrac{7}{4\lambda}
\ddot{u}_x\right)-\right.\\
&-\left.r^5_{j0}\dfrac{3}{\lambda}\ddot{f}-\dot{r}_{j0}\left(\dfrac{8}{\lambda}
r^4_{j0}\dot{f}+\dfrac{1}{\lambda}r^2_{j0}\dot{u}_x\right)+\dot{r}_{j0}^2
\left(-\dfrac{3}{\lambda}r_{j0}^3\,f+\dfrac{3}{2\lambda}r_{j0}u_x\right)+\right.\\
&\left.+\ddot{r}_{j0}\left(-\dfrac{7}{4\lambda}r^4_{j0}\,f-\dfrac{1}{4\lambda}
r^2_{j0}\,u_x\right)\right]+\dfrac{\partial^2 r_{j1}}{\partial z^2}
\left[\dfrac{b_{c2}}{\Ca}-r^3_{j0}\dfrac{1}{2\lambda}\dot{u}_x-
r^5_{j0}\dfrac{1}{\lambda}\dot{f}-\right.\\
&\left.-\dot{r}_{j0}\left(\dfrac{7}{4\lambda}r^4_{j0}\,f+\dfrac{1}{4\lambda}
u_x\,r^2_{j0}\right)\right]+\dfrac{\partial^3 r_{j1}}{\partial z^3}
\left[\dfrac{b_{c3}}{\Ca}-\dfrac{1}{4\lambda}u_x\,r^3_{j0}-\dfrac{1}{4\lambda}r^5_{j0}\,f\right]+\\
&+\dfrac{\partial^4 r_{j1}}{\partial z^4}\dfrac{b_{c4}}{\Ca}=0\,.\label{Continuidad2}
\end{split}
\end{equation}
Notice that the five functions $b_{ci}$ in equation~\eqref{Continuidad2} are deduced from the
fact that
\begin{equation}
\begin{split}
&-\dfrac{1}{8\lambda\Ca}\dfrac{\partial}{\partial z}\left(r^4_j\,\dot{\mathcal{C}}\right)=
-\dfrac{3}{2\lambda\Ca}r_{j0}^{-2}\dot{r}_{j0}\,\dot{\mathcal{C}}_0\,r_{j1}-
\dfrac{1}{2\lambda\Ca}r^3_{j0}\,\dot{\mathcal{C}}_0\dfrac{\partial r_{j1}}{\partial z}-
\dfrac{1}{2\lambda\Ca}r^3_{j0}\,\ddot{\mathcal{C}}_0\,r_{j1}-\\
&-\dfrac{1}{2\lambda\Ca}r^3_{j0}\dot{r}_{j0}\,\dot{\mathcal{C}}_1-
\dfrac{1}{8\lambda\Ca}r^4_{j0}\,\ddot{\mathcal{C}}_1\,,
\end{split}
\end{equation}
where
\begin{equation}
\dot{\mathcal{C}}_0=\frac{{\rm d}\mathcal{C}_0}{{\rm d}z}=
-\dot{r}_{j0}\,r_{j0}^{-2}d_0^{-1/2}-\dot{r}_{j0}\ddot{r}_{j0}\,r_{j0}^{-1}d_0^{-3/2}+
3\dot{r}_{j0}\ddot{r}_{j0}^2\,d_0^{-5/2}-r^{(3)}_{j0}d_0^{-3/2}\,,
\end{equation}
and
\begin{equation}
\begin{split}
&\ddot{\mathcal{C}}_0=\frac{{\rm d}^2\mathcal{C}_0}{d z^2}=
2\dot{r}_{j0}^2\,r_{j0}^{-3}d_0^{-1/2}+2\dot{r}_{j0}^2\ddot{r}_{j0}\,r_{j0}^{-2}d_0^{-3/2}-
\ddot{r}_{j0}r_{j0}^{-2}\,d_0^{-1/2}+3\dot{r}_{j0}^2\ddot{r}_{j0}^2r_{j0}^{-1}d_0^{-5/2}-\\
&-\ddot{r}_{j0}^2\,r_{j0}^{-1}d_0^{-3/2}-15\dot{r}_{j0}^2\ddot{r}_{j0}^3d_0^{-7/2}+
3\ddot{r}_{j0}^3\,d_0^{-5/2}-\dot{r}_{j0}r^{(3)}_{j0}r^{-1}_{j0}d_0^{-3/2}+
9\dot{r}_{j0}\ddot{r}_{j0}r^{(3)}_{j0}\,d_0^{-5/2}-\\&-r^{(4)}\,d_0^{-3/2}\,,
\end{split}
\end{equation}
with $d_0=1+\dot{r}_{j0}^2$.
\\

Therefore,
\begin{equation}
\begin{split}
&b_{c0}=-\dfrac{3}{2\lambda}r^2_{j0}\dot{r}_{j0}\,\dot{\mathcal{C}}_0-
\dfrac{1}{2\lambda}r^3_{j0}\,\ddot{\mathcal{C}}_0-\dfrac{1}{4\lambda}\dot{r}_{j0}^2d_0^{-1/2}-\\
&-\dfrac{1}{8\lambda}\left(2\ddot{r}_{j0}r_{j0}d_0^{-1/2}-
3r^2_{j0}\dot{r}_{j0}^2\ddot{r}_{j0}^2\,d_0^{-5/2}+\ddot{r}^2_{j0}r^2_{j0}\,d_0^{-3/2}+
\dot{r}_{j0}r^{(3)}_{j0}r^2_{j0}\,d_0^{-3/2}\right)\,,\label{bc0}
\end{split}
\end{equation}
\begin{equation}
\begin{split}
&b_{c1}=-\dfrac{1}{2\lambda}r^3_{j0}\,\dot{\mathcal{C}}_0-
\dfrac{1}{4\lambda}r_{j0}\dot{r}_{j0}^3\,d_0^{-3/2}-\dfrac{3}{4\lambda}r_{j0}^2\dot{r}_{j0}^3
\ddot{r}_{j0}d_0^{-5/2}-\dfrac{1}{8\lambda}r_{j0}^2\dot{r}_{j0}\ddot{r}_{j0}d_0^{-3/2}-\\
&-\dfrac{21}{8\lambda}r^3_{j0}\dot{r}_{j0}\ddot{r}_{j0}^2\,d_0^{-5/2}+
\dfrac{75}{8\lambda}r^3_{j0}\dot{r}_{j0}^3\ddot{r}_{j0}^2\,d_0^{-7/2}-
\dfrac{15}{8\lambda}r^3_{j0}r^{(3)}_{j0}\dot{r}_{j0}^2d_0^{-5/2}-
\dfrac{105}{8\lambda}r^4_{j0}\dot{r}_{j0}^3\ddot{r}_{j0}^3\,d^{-9/2}_0+\\
&+\dfrac{45}{8\lambda}r^4_{j0}\dot{r}_{j0}\ddot{r}^3_{j0}\,d_0^{-7/2}+
\dfrac{45}{8\lambda}r_{j0}^4\dot{r}_{j0}^2\ddot{r}_{j0}r^{(3)}_{j0}d_0^{-7/2}-
\dfrac{9}{8\lambda}r^4_{j0}\ddot{r}_{j0}r^{(3)}_{j0}d^{-5/2}_0-\\
&-\dfrac{3}{8\lambda}r^4_{j0}r^{(4)}_{j0}\dot{r}_{j0}\,d_0^{-5/2}+
\dfrac{1}{8\lambda}\,r^{(3)}_{j0}r^3_0\,d_0^{-3/2},\label{bc1}
\end{split}
\end{equation}
\begin{equation}
\begin{split}
&b_{c2}=\dfrac{1}{4\lambda}r^2_{j0}\dot{r}_{j0}^2d_0^{-3/2}-
\dfrac{30}{8\lambda}r_{j0}^3\dot{r}_{j0}^2\ddot{r}_{j0}d_0^{-5/2}+
\dfrac{1}{8\lambda}\,r_{j0}^{2}d_0^{-1/2}+\dfrac{1}{4\lambda}r^3_{j0}\ddot{r}_{j0}d_0^{-3/2}+\\
&+\dfrac{45}{8\lambda}\,r^4_{j0}\dot{r}_{j0}^2\ddot{r}_{j0}^2d_0^{-7/2}-
\dfrac{9}{8\lambda}r^4_{j0}\ddot{r}_{j0}^2d_0^{-5/2}-
\dfrac{9}{8\lambda}r^4_{j0}\dot{r}_{j0}r^{(3)}_{j0}\,d_0^{-5/2}\,,\label{bc2}
\end{split}
\end{equation}
\begin{equation}
\begin{split}
&b_{c3}=\dfrac{5}{8\lambda}\,r_{j0}^3\dot{r}_{j0}\,d_0^{-3/2}-
\dfrac{9}{8\lambda}\,r^4_{j0}\dot{r}_{j0}\ddot{r}_{j0}\,d_0^{-5/2}\,,\label{bc3}
\end{split}
\end{equation}
and
\begin{equation}
b_{c4}=\dfrac{1}{8\lambda}r^4_{j0}d_0^{-3/2}\,.\label{bc4}
\end{equation}
\\

Equation~\eqref{Continuidad2} admits solutions of the form
$r_{j1}(z,t)=e^{\omega t}\,\bar{r}_{j1}(z)$ leading, through the Chebychev discretization
explained in the main text, to the following linear system of equations for the $N$ eigenvalues
and their corresponding eigenfunctions,
\begin{equation}
\omega\mathcal{A}\cdot\mathbf{\bar{r}_{j1}}=\mathcal{B}\cdot\mathbf{\bar{r}_{j1}}\,,\label{eceig21}
\end{equation}
with $\mathbf{\bar{r}_{j1}}$ denoting a column vector containing the values of the perturbed
radius at the $N$ collocation points,
\begin{equation}
\mathcal{A}=\mathrm{diag}\left(-\frac{1}{4\lambda}\,r^3_{j0}\right)\mathcal{D}^2-
\mathrm{diag}\left(\frac{1}{2\lambda}\,r^2_{j0}\dot{r}_{j0}\right)\mathcal{D}+
\mathrm{diag}\left(2\,r_{j0}+\frac{1}{4\lambda}\left(2\,r_{j0}\dot{r}_{j0}^2+r^2_{j0}\,
\ddot{r}_{j0}\right)\right)\mathcal{I}\,,\label{AA}
\end{equation}
and
\begin{equation}
\mathcal{B}=\mathrm{diag}\left(B_4\right)\mathcal{D}^4+\mathrm{diag}\left(B_3\right)\mathcal{D}^3+
\mathrm{diag}\left(B_2\right)\mathcal{D}^2+\mathrm{diag}\left(B_1\right)\mathcal{D}+\mathrm{diag}\left(B_0\right)\, ,
\end{equation}
with the five functions $B_i$ given by
\begin{equation}
\begin{split}
&B_0=\dfrac{b_{c0}}{\Ca}+2r_{j0}\dot{u}_x+4r^3_{j0}\dot{f}-
r^3_{j0}\left(\dfrac{3}{2\lambda}u^{(3)}_x+\dfrac{2}{\lambda}\dot{f}\right)-
r^5_{j0}\dfrac{9}{4\lambda}f^{(3)}+\\
&+\dot{r}_{j0}\left(2\,u_x-\dfrac{21}{4\lambda}r^2_{j0}\ddot{u}_x+
fr^2_{j0}\left(12-\dfrac{6}{\lambda}\right)-\dfrac{15}{\lambda}r^4_{j0}\ddot{f}\right)+
\dot{r}_{j0}^2\left(-\dfrac{16}{\lambda}r^3_{j0}\dot{f}-
\dfrac{1}{\lambda}r_{j0}\dot{u}_x\right)+\\
&+\dot{r}_{j0}^3\left(-\dfrac{3}{\lambda}r^2_{j0}\,f+\dfrac{1}{2\lambda}u_x\right)-
\ddot{r}_{j0}\left(\dfrac{5}{\lambda}r^4_{j0}\dot{f}+\dfrac{3}{2\lambda}r^2_{j0}\dot{u}_x\right)-
\dot{r}_{j0}\ddot{r}_{j0}\left(\dfrac{7}{\lambda}r^3_{j0}\,f+\dfrac{1}{2\lambda}u_x\,r_{j0}\right)-\\
&-r^{(3)}_{j0}\left(\dfrac{3}{4\lambda}u_xr^2_{j0}+\dfrac{5}{4\lambda}r^4_{j0}\,f\right)\,,
\label{AB0}
\end{split}
\end{equation}
\begin{equation}
\begin{split}
&B_1=\dfrac{b_{c1}}{\Ca}+2r_{j0}u_x+r^{3}_{j0}\left(f\left(4-\dfrac{2}{\lambda}\right)-
\dfrac{7}{4\lambda}\ddot{u}_x\right)-r^5_{j0}\dfrac{3}{\lambda}\ddot{f}-
\dot{r}_{j0}\left(\dfrac{8}{\lambda}r^4_{j0}\dot{f}+\dfrac{1}{\lambda}r^2_{j0}\dot{u}_x\right)+\\
&+\dot{r}_{j0}^2\left(-\dfrac{3}{\lambda}r_{j0}^3\,f+\dfrac{3}{2\lambda}r_{j0}u_x\right)+
\ddot{r}_{j0}\left(-\dfrac{7}{4\lambda}r^4_{j0}\,f-\dfrac{1}{4\lambda}r^2_{j0}\,u_x\right)\,,
\label{AB1}
\end{split}
\end{equation}
\begin{equation}
B_2=\dfrac{b_{c2}}{\Ca}-r^3_{j0}\dfrac{1}{2\lambda}\dot{u}_x-r^5_{j0}\dfrac{1}{\lambda}\dot{f}-
\dot{r}_{j0}\left(\dfrac{7}{4\lambda}r^4_{j0}\,f+\dfrac{1}{4\lambda}u_x\,r^2_{j0}\right)\,,
\label{AB2}
\end{equation}
\begin{equation}
B_3=\dfrac{b_{c3}}{\Ca}-\dfrac{1}{4\lambda}u_x\,r^3_{j0}-\dfrac{1}{4\lambda}r^5_{j0}\,f\,,
\label{AB3}
\end{equation}
and
\begin{equation}
B_4=\frac{1}{8\lambda\Ca}r^4_{j0}d_0^{-3/2}\,,\label{AB4}
\end{equation}
with $b_{ci}$ given by equations (\ref{bc0})-(\ref{bc4}).

\bibliographystyle{jfm}

%\bibliography{pinchoff2}

\end{document}